\documentclass[a4paper,fleqn, review]{cas-sc}

\usepackage[numbers]{natbib}
\usepackage{mathtools}
\usepackage{float}
\usepackage{algorithm}
\usepackage{algpseudocode}
\usepackage{siunitx}
\usepackage{pgfplots}
\usepackage{setspace}
\usepackage[mathlines]{lineno}
\def\tsc#1{\csdef{#1}{\textsc{\lowercase{#1}}\xspace}}
\tsc{WGM}
\tsc{QE}
\newcommand{\bs}[1]{\boldsymbol{#1}}

\newcommand{\mdl}{\mathcal{M}}
\newcommand{\surmdl}[1]{\widetilde{\mathcal{M}}_{#1}}

\newcommand{\rndinp}{\boldsymbol{\xi}}
\newcommand{\rndinps}{\xi}

\newcommand{\rnddsn}{\boldsymbol{\delta}}
\newcommand{\rnddsns}{\delta}
\newcommand{\jrndinp}{\boldsymbol{\zeta}}

\newcommand{\auginp}{\boldsymbol{z}}
\newcommand{\auginps}{z}

\newcommand{\rndouts}{Y}
\newcommand{\relouts}{y}

\newcommand{\inp}{\boldsymbol{x}}
\newcommand{\inps}{x}

\newcommand{\augDataset}{\mathcal{D}}

\newcommand{\din}{d_{\inp}}
\newcommand{\stochin}{d_{\rndinp}}
\newcommand{\quant}{q}

\newcommand{\dsnSpace}{\mathcal{X}}
\newcommand{\R}{\mathbb{R}}
\newcommand{\augDomain}{\mathcal{Z}}

\newcommand{\hyperpara}{\boldsymbol{\lambda}}
\newcommand{\dataset}{\mathcal{D}}
\newcommand{\designMat}{\mathbf{Z}}
\newcommand{\outvec}{\mathbf{y}}
\newcommand{\GP}{\mathcal{GP}}
\newcommand{\meaneval}{\mathbf{m}}

\renewcommand{\P}{\mathbb{P}}
\newcommand{\E}{\mathbb{E}}
\newcommand{\V}{\mathbb{V}}
\newcommand{\id}{\mathbb{I}}

\DeclareMathOperator*{\argmin}{arg\,min}
\DeclareMathOperator*{\argmax}{arg\,max}
\DeclarePairedDelimiter{\ceil}{\lceil}{\rceil}

\newcommand{\cdf}[1]{F_{#1}}
\newcommand{\pdf}[1]{p_{#1}}

\begin{document}
%\linenumbers
\let\WriteBookmarks\relax
\def\floatpagepagefraction{1}
\def\textpagefraction{.001}

% Short title
\shorttitle{}    

% Short author
\shortauthors{}  

% Main title of the paper
\title [mode = title]{Efficient Robust Optimization via an Adaptive Mono-Level Gaussian Process Surrogate}  

% Title footnote mark
% eg: \tnotemark[1]
% \tnotemark[1] 

% Title footnote 1.
% eg: \tnotetext[1]{Title footnote text}
% \tnotetext[1]{} 

% First author
%
% Options: Use if required
% eg: \author[1,3]{Author Name}[type=editor,
%       style=chinese,
%       auid=000,
%       bioid=1,
%       prefix=Sir,
%       orcid=0000-0000-0000-0000,
%       facebook=<facebook id>,
%       twitter=<twitter id>,
%       linkedin=<linkedin id>,
%       gplus=<gplus id>]

\author[1]{Julius Schultz}[orcid=0000-0002-1486-9976]
% Corresponding author indication
\cormark[1]
% Footnote of the first author
% \fnmark[1]
% Email id of the first author
\ead{j.schultz@tu-braunschweig.de}
% URL of the first author
% \ead[url]{}
% Credit authorship
% eg: \credit{Conceptualization of this study, Methodology, Software}
\credit{Conceptualization, Data curation, Methodology, Software, Formal analysis, Validation, Visualization, Writing - original draft, Writing - review \& editing}

% Address/affiliation
\affiliation[1]{organization={Technische Universität Braunschweig},
            addressline={Institute for Acoustic and Dynamics}, 
            city={Braunschweig},
%          citysep={}, % Uncomment if no comma needed between city and postcode
            postcode={38106}, 
            % state={},
            country={Germany}}

\author[1]{Mustafa Essou}[orcid=0009-0005-8676-8823]
% Email id of the second author
\ead{mustafa.essou@tu-braunschweig.de}
% URL of the second author
% \ead[url]{}
% Credit authorship
\credit{Investigation, Writing - review \& editing}

\author[1]{Ulrich Römer}[orcid=0000-0002-1277-7509]
% Email id of the second author
\ead{u.roemer@tu-braunschweig.de}
% URL of the second author
% \ead[url]{}
% Credit authorship
\credit{Conceptualization, Supervision, Funding acquisition, Methodology, Writing - review \& editing}

% Corresponding author text
\cortext[1]{Corresponding author}
% Footnote text
% \fntext[1]{}

% For a title note without a number/mark
%\nonumnote{}

% Here goes the abstract
\begin{abstract}
Optimization under uncertainty seeks designs that perform robustly despite varying operating conditions or uncertain model parameters. However, identifying such designs typically requires solving an optimization problem with a nested uncertainty analysis, resulting in a high computational burden. Surrogate models can alleviate this burden and enable optimization under uncertainty even with computationally expensive simulation models. This paper proposes a novel surrogate-based approach that reduces the number of costly model evaluations required for robust optimization. The approach adopts a mono-level strategy, in which a single Gaussian process surrogate is constructed and adaptively refined for both optimization and uncertainty analysis. Refinement in the design space is guided by a sampling-based uncertainty indicator that propagates the surrogate's epistemic uncertainty to the robustness measure. Efficient uncertainty estimation is crucial for the scalability of the algorithm and achieved through pathwise sampling combined with a weight-space representation of the Gaussian process. The proposed approach is evaluated on several analytical benchmark models and requires fewer model evaluations than a state-of-the-art bi-level surrogate approach while achieving comparable solution accuracy. Once trained, the mono-level surrogate can be reused to optimize different robustness measures at reduced cost. Moreover, the approach is successfully applied to a structural optimization problem with constraints and design variable uncertainty. The code is made available on \href{https://github.com/JS273/Robust-Optimization-via-a-Mono-Level-Gaussian-Process-Surrogate}{GitHub} and \href{https://zenodo.org/records/24201161}{Zenodo}.
\end{abstract}

% Use if graphical abstract is present
%\begin{graphicalabstract}
%\includegraphics{}
%\end{graphicalabstract}

% Research highlights
% \begin{highlights}
% \item Data-efficient robust optimization with a mono-level Gaussian process surrogate 
% \item Efficient posterior sampling to propagate epistemic surrogate uncertainty
% \item Validated on analytical benchmarks and a structural optimization problem
% \item Outperforms a bi-level surrogate-based robust optimization approach
% \item Reuses model evaluations across optimization runs
% \end{highlights}

% Keywords
% Each keyword is seperated by \sep

\maketitle

% Main text
\section{Introduction}
% The text width of this document is: \the\textwidth. \\

Design optimization is an essential aspect of many engineering disciplines to maximize the performance of a system. Classical deterministic design optimization assumes nominal conditions and may therefore yield designs with poor performance or even failure under off-design conditions or changing environments. This motivates the explicit treatment of uncertainties, e.g., from fluctuations in environmental parameters, manufacturing tolerances, material variability, and changing load conditions. Incorporating these uncertainties into the optimization process enables the identification of robust and reliable designs. 

The general field of optimization under uncertainty can be divided into two main branches that have been developed over the past decades \citep{schueller2008computational}. Within \textit{reliability-based design optimization} (RBDO) \citep{enevoldsen1994reliability, gasser1997reliability}, a cost function is minimized while ensuring that its probability of failure is below a prescribed threshold. 
The second branch is named \textit{robust design optimization} (RDO) \citep{beyer2007robust, mulvey1995robust, park2006robust} and aims to optimize a system's performance that is subject to uncertainty. In this case the performance measure is stochastic and robustness measures are required to guide the optimization process. Typical robustness measures are statistical moments \citep{mulvey1995robust}, quantiles \citep{du2004integrated} or probability of failure. Taking a quantile as a robustness measure will lead to designs that guarantee a certain performance level with a specified probability. Optimizing a weighted sum of mean and standard deviation balances expected performance against sensitivity to uncertainty. In this work we will focus on RDO problems. Please note that robust optimization, in its classical form, employs a set-based worst-case formulation, see \citet{ben1998robust}. Here, we use RDO following \citet{beyer2007robust}, as an approach employing a probabilistic robustness measure.

% Literature review 
% - Application of RDO
RDO has been applied in many different engineering fields. The applications range from structural mechanics \citep{doltsinis2004robust, ni2020reliability} and automotive engineering \citep{hwang2001robust, fang2015multiobjective} to aerodynamics \citep{cook2018effective, sabater2022development, sabater2020efficient}. 

% - Approximation techiques (FOSM)
% Challenge --> Computational burden
A major challenge for RDO problems is their computational cost. During optimization, the design space must be explored, while each candidate design requires an uncertainty analysis to estimate the robustness measure. This nested structure of many-query tasks can become prohibitively expensive when each evaluation of the underlying performance model requires a costly computer simulation. Consequently, a range of surrogate-based RDO strategies have been proposed in the literature as summarized below. 

Several surrogate-based RDO methods construct a surrogate for the mapping from the design space to the robustness measure and subsequently optimize this surrogate over the design space. These methods differ mainly in how the robustness measure is determined at training points. \citet{nan2024gradient} use a direct Monte Carlo (MC) simulation, which can require many expensive model evaluations. The works of \citet{keane2012cokriging} and \citet{keane2020robust} employ a multi-fidelity approach, that combines many coarse and some high-fidelity estimates of the robustness measure in a co-kriging surrogate. The approaches also differ in their adaptive refinement strategies. A purely exploitation-based approach is used in \citet{keane2020robust} whereas \citet{nan2024gradient} additionally incorporate the surrogate uncertainty to enable exploration.
For high-dimensional stochastic inputs, quantile regression approaches have been introduced in \citet{sabater2021bayesian} and \citet{torossian2020review}. These approaches avoid an explicit representation of the stochastic model behavior and are thus insensitive to the curse of dimensionality.  

% Emerging methods for adaptive UQ and Opt --> Bayesian opt --> Bi-level approaches
Bayesian optimization is another widely used approach for efficient global optimization of expensive functions \cite{jones1998efficient}. It uses acquisition functions derived from posterior surrogate predictions to decide where to adaptively refine a Gaussian process (GP) surrogate. The works of \citet{razaaly2020quantile, sabater2020efficient} employ Bayesian optimization for the design optimization task and combine it with surrogate-based uncertainty quantification to estimate the robustness measure. They construct an outer surrogate in the design space and local surrogates to emulate the stochastic model behavior. These methods are therefore referred to as \textit{bi-level} approaches.  

% Drawback of bi-level approaches --> 
A limitation of the bi-level approach is that new local surrogate models are constructed from scratch at each candidate design point, neglecting available information on neighboring local surrogates. Information reuse strategies have been proposed to improve the data efficiency for local surrogate construction in \citet{essou2024}. Mono-level approaches seek to further minimize the required number of model evaluations by using a single surrogate for both optimization and uncertainty quantification. This surrogate is constructed over the joint space of design variables and stochastic parameters. Although this augmented space can be high-dimensional, accurate modeling is primarily required in regions that are relevant to the robust optimum. 

% Mono-level approaches
Several mono-level approaches have been proposed. In \citet{archbold2024variational} a mono-level surrogate is constructed based on Bayesian principles, such that the posterior accounts directly for both input parameter uncertainty and epistemic uncertainty due to limited training data. This method incorporates dimensionality reduction, but does not focus on adaptive sampling strategies for data-efficient surrogate refinement.
An adaptive approach has been proposed in \citet{zhang2017sequential}. Therein, a mono-level surrogate is constructed in a sub-region of the design space and this sub-region is shifted if the optimum of the current iteration reaches a local sub-region boundary. However, the local approach restricts the search to a neighborhood of the current optimum, which may hinder global exploration. 

\citet{janusevskis2013simultaneous} and \citet{beland2017bayesian} build a mono-level GP surrogate and use the fact that GPs are closed under linear operators. Integrating over the uncertain inputs therefore gives a GP of the mean response directly, and Bayesian optimization acquisition functions can be applied to it. In \citet{janusevskis2013simultaneous}, the integrals are evaluated in closed form, which requires a Gaussian kernel and Gaussian inputs, while \citet{beland2017bayesian} extend the approach to design-variable uncertainty and probabilistic constraints, and use separable kernels or sparse quadrature to evaluate the integrals. Both approaches are limited to linear robustness measures.

In the context of RBDO, \citet{moustapha2016quantile} use a single GP surrogate in an augmented design space \citep{dubourg2011reliability} to estimate quantiles defining probabilistic constraints, and refine it with the U-function criterion \citep{echard2011ak} with respect to the fixed constraint threshold. In contrast, in this work the quantile is the objective itself, so its target value is unknown and the surrogate refinement is guided by the posterior distribution of the quantile.

The work most closely related to this paper is \citet{cakmak2020bayesian}, who optimize the value-at-risk, i.e., a quantile, and the conditional value-at-risk using a mono-level GP surrogate. The next evaluation point in the joint space is selected by a one-step Bayes-optimal knowledge-gradient acquisition function. However, evaluating this acquisition function requires solving a nested optimization over fantasy models \citep{snoek2012practical}, and the robustness measure is estimated from posterior samples drawn via Cholesky factorization, whose cost grows cubically with the number of uncertainty samples. Both factors restrict the approach to a small number of uncertainty samples and therefore to coarse estimates of the robustness measure. Furthermore, uncertainties are restricted to environmental variables that are independent of the design. In contrast, the approach proposed in this paper relies on cheaper acquisition functions and an efficient posterior sampling scheme that scales linearly with the number of samples, enabling accurate quantile estimates.

% Contribution
To sum up, in this paper, we propose a mono-level approach for RDO that jointly targets nonlinear robustness measures such as quantiles with design-variable uncertainty and probabilistic constraints, scales to large sample sizes for accurate quantile estimation, and allows information reuse across different robustness measures.
To enable adaptive sampling, we propagate the epistemic uncertainty of the mono-level surrogate to the robustness measure. Every update of the mono-level surrogate results in an updated uncertainty estimate of the robustness measure, enabling efficient exploration of the design space.
To enable efficient posterior sampling, we leverage a finite-dimensional weight-space representation of GPs together with pathwise sampling, allowing scalability of the proposed method. The approach can handle uncertainty in design variables as well as model parameters and has been tested for up to 9 dimensional problems. The performance of the approach is demonstrated on common analytical benchmark functions and on a finite-element structural optimization benchmark.

% Strucutre of the paper
The remainder of this paper is organized as follows: Section \ref{sec:ProbFormulation} formulates the RDO problem and Section \ref{sec:GPregression} reviews the main steps of GP regression. Section \ref{sec:MonoLevel} introduces the proposed mono-level robust optimization approach, including the algorithm, acquisition policies, efficient posterior sampling, and an illustrative 1D example. Section \ref{sec:App} presents results for analytical benchmarks and a finite-element structural optimization problem. Finally, Section \ref{sec:Discussion} compares the performance of the mono-level approach against a state-of-the-art bi-level approach and discusses its flexibility and scalability.

\section{Problem formulation}
\label{sec:ProbFormulation}
% The text width of this document is: \the\textwidth.
Throughout this work, we consider an expensive computational model $\mdl(\inp + \rnddsn(\omega), \rndinp(\omega)) \in \mathbb{R}$, where $\omega \in \Omega$ denotes an elementary random outcome associated to the probability space $(\Omega, \mathcal{B}_{\Omega}, \P_{\Omega})$. The vector $\inp \in \dsnSpace \subset \R^{\din}$ contains the controllable design variables, while deviations from the nominal design are represented by the random vector $\rnddsn : \Omega \rightarrow \widetilde{\mathcal{X}} \subset \R^{\din}$ with $\E_{\P_{\Omega}}[\rnddsn] = \boldsymbol{0}$. Such deviations could arise for example from manufacturing imperfections. Uncertainties in model parameters, describing changing load, boundary, or environmental conditions, are represented by the random vector $\rndinp : \Omega \to \mathcal{S} \subseteq \R^{\stochin}$. 

For compact notation, we collect both uncertainty sources in the joint random vector $\jrndinp := (\rnddsn, \rndinp)$, with $\jrndinp : \Omega \rightarrow \mathcal{Q}$ and $\mathcal{Q} = \widetilde{\mathcal{X}} \times \mathcal{S}$. Moreover, we introduce the augmented design space $\augDomain = \dsnSpace' \, \times \, \mathcal{S}'$ which has dimension $d_z = d_\xi + d_x$.
This augmented design space is composed of box-domains $\dsnSpace', \mathcal{S}'$ associated with the design variables and uncertain model inputs. 
The design space $\dsnSpace$ is extended due to the influence of $\rnddsn$ yielding
\begin{equation}
    \dsnSpace' = \prod_{i = 1}^{\din} [x_i^l + \rnddsns_i^l, x_i^u + \rnddsns_i^u]
\end{equation}
where $\rnddsns_i^l$ and $\rnddsns_i^u$ denote the lower and upper bound for the $i$th component of the design perturbation. For continuous random variables with unbounded support, these lower and upper bounds are defined as marginal tail quantiles, i.e., $\rnddsns_i^l = \cdf{\rnddsn}^{-1}(0.01)$ and $\rnddsns_i^u = \cdf{\rnddsn}^{-1}(0.99)$. 
Analogously, we define a box-domain associated with the random vector $\rndinp$ by
\begin{equation}
    \mathcal{S}' = \prod_{i = 1}^{\stochin} [\rndinps_i^l, \rndinps_i^u].
\end{equation}
For a fixed design $\inp$, the random performance measure is defined by 
\begin{equation}
    \rndouts_{\inp}(\omega)
    := \mdl(\inp + \rnddsn(\omega), \rndinp(\omega)) 
\end{equation}
We seek a design that minimizes a robustness measure $\mathcal{J}: \dsnSpace \rightarrow \R$ while satisfying probabilistic feasibility constraints  
\begin{align}
    \inp^{*} = \argmin_{\inp \in \dsnSpace} \mathcal{J}(\inp) \quad \quad s.t. \quad \quad \P_{\Omega}[c_j(\inp + \rnddsn) \leq 0] \geq 1 - \eta , \ j = 1, \dots , d_c,
    \label{eq:robOptProb}
\end{align}
where $c_j: \dsnSpace' \rightarrow \R$ denotes the $j$th constraint function and $\eta \in [0, 1]$ is the prescribed allowed probability of constraint violation.

% Bayes risk
A robustness measure is commonly defined in terms of statistics of the random performance measure. A widely used approach is a weighted sum of the mean and standard deviation \citep{lee2001robust, mulvey1995robust}, i.e.,
\begin{align}
    \mathcal{J}(\inp) = \frac{(1 - \alpha)}{\bar{\mu}} \E_{\P_{\Omega}}[\mdl(\inp + \rnddsn, \rndinp)] + \frac{\alpha}{{\bar{\sigma}}} \sqrt{\V_{\P_{\Omega}}[\mdl(\inp + \rnddsn, \rndinp)]} \label{eq:mean_std_meas}
\end{align}
where $\bar{\mu} \in \R$ and $\bar{\sigma} \in \R$ denote normalization constants.
Optimizing this robustness measure yields a design that aims at good performance with low variability with respect to the joint uncertainty in $\jrndinp$. The trade-off between mean performance and low variability is controlled by the choice of $\alpha \in [0,\,1]$. \\
In safety-critical applications it is often desirable to guarantee a certain level of performance with prescribed probability. For a minimization problem, a high $\tau$-quantile 
\begin{align}
    \mathcal{J}(\inp \vert \tau) = F^{-1}_{\rndouts_{\inp}}(\tau) := \inf\{ \relouts \mid F_{\rndouts_{\inp}}(\relouts) \geq \tau\} \label{eq:quant_def}
\end{align}
serves as a conservative robustness measure \citep{razaaly2020quantile, sabater2020efficient, quagliarella2014optimization}, where $F_{\rndouts_{\inp}}$ denotes the cumulative distribution function (CDF) of the random performance measure $\rndouts_{\inp}$ at design $\inp$.

A $\tau$-quantile $q_{\tau}$ is a cut-off point in the range of a random variable. For a continuous random variable, it satisfies $\P_{\Omega}[\rndouts_{\inp} < q_{\tau}] = \tau$. Formally, it is defined via the inverse CDF as given in \eqref{eq:quant_def}. In practice the quantile is often estimated using MC sampling \citep{van2000asymptotic}. 
To this end, independent and identically distributed realizations $\relouts_i = \rndouts_{\inp}(\omega_i), i = 1,\ldots,n$, are drawn and sorted to obtain the order statistics $\relouts_{(1)} \leq \relouts_{(2)} \leq \cdots \leq \relouts_{(n)}$. The MC estimator of the $\tau$-quantile is then defined as 
\begin{align}
    q_{\tau} \approx q_{\tau}^{n} := \relouts_{(\ceil{n \tau})}, \qquad \tau \in (0,1),
\end{align}
where $\ceil{n \tau}$ denotes the smallest integer greater than or equal
to $n\tau$. Alternative approaches include fitting a parametric distribution to the data \citep{noufaily2013parametric} or expressing quantile estimation as the minimization of an asymmetric linear loss \citep{Koenker_2005}. In this work, we use the MC estimator and leverage surrogate models to reduce its computational cost.

\section{Gaussian process regression}
\label{sec:GPregression}
Gaussian process regression \citep{gramacy2020surrogates, williams2006gaussian} is a probabilistic surrogate modeling technique that is widely used to speed up many-query tasks like optimization \citep{forrester2008engineering} or uncertainty quantification \citep{owen2017comparison}. 
The starting point is a GP prior constructed over the augmented design space
\begin{align}
    f \sim \GP(m(\auginp), k(\auginp, \auginp'))
\end{align}
where $m: \R^{d_z} \rightarrow \R$ is the mean function and $k: \R^{d_z} \, \times \, \R^{d_z} \rightarrow \R$ the covariance function. The mean function specifies the prior trend, whereas the covariance (kernel) function allows one to express assumptions on the target function, such as its smoothness, characteristic length scales or periodicity properties. A popular choice is the Mat\'ern kernel \citep{williams2006gaussian}
\begin{align}
    k\!\left(\auginp, \auginp'\right)
    &=
    \sigma_f^2
    \frac{2^{1-\nu}}{\Gamma(\nu)}
    \left(
        \sqrt{2\nu}\,r(\auginp,\auginp')
    \right)^{\nu}
    K_{\nu}\!\left(
        \sqrt{2\nu}\,r(\auginp,\auginp')
    \right), \quad \text{with}
    \\
    r(\auginp,\auginp')
    &=
    \left[
        \sum_{i=1}^{d_z}
        \left(
            \frac{\auginps_i-\auginps_i'}{\ell_i}
        \right)^2
    \right]^{1/2}.
\end{align}
Here, $K_\nu$ denotes the modified Bessel function, 
$\sigma_f^2$ is the process variance and
$\boldsymbol{\ell}=(\ell_1,\ldots,\ell_{d_z})^\top$ contains the
dimension-specific length-scales. The parameter $\nu>0$ controls the smoothness of the GP sample paths and is set to $5/2$ in this work. \\
% Observation data
Given a set of observation data $\dataset = \{(\auginp^{(i)}, \relouts^{(i)}) \mid i = 1, \dots, n_t\}$ we can now define the design matrix $\designMat \in \R^{d_z \times n_t}$ and the observation vector $\outvec \in \R^{n_t}$. The outputs $\relouts^{(i)}$ correspond to noisy observations of the target function $\relouts^{(i)} = f(\auginp^{(i)}) + \varepsilon^{(i)}$ with $\varepsilon^{(i)} \overset{\mathrm{i.i.d.}}{\sim} \mathcal{N}(0,\sigma_n^2)$, where $\sigma_n^2$ denotes the homoscedastic observation-noise variance. In this work we deal with simulated training data so we set $\sigma_n^2 = 10^{-4}$ to a small value for numerical stability. Conditioning the GP prior on $\dataset$ yields the posterior GP
% Posterior
\begin{align}
    f_{\dataset} \sim \GP(m_{\dataset}(\auginp), k_{\dataset}(\auginp, \auginp')).
\end{align}
Let $ \mathbf{K} := K(\designMat,\designMat) \in \R^{n_t \times n_t}$ denote the covariance matrix and $\meaneval := \left[m(\auginp^{(1)}),\ldots,m(\auginp^{(n_t)})\right]^\top \in \R^{n_t}$ the mean vector evaluated at the training inputs. The posterior mean and covariance function are given by
\begin{align}
    m_{\dataset}(\auginp) &= m(\auginp) + K(\auginp, \designMat) (\mathbf{K} + \sigma_n^2 \id_{n_t})^{-1} (\outvec - \meaneval),\\
    k_{\dataset}(\auginp, \auginp') &= k(\auginp, \auginp') - K(\auginp, \designMat) (\mathbf{K} + \sigma_n^2 \id_{n_t})^{-1} K(\designMat, \auginp'),
\end{align}
where $K(\auginp, \designMat) \in \R^{1 \times n_t}$ is a vector of cross covariances with entries $k(\auginp,\auginp^{(i)})$ and likewise $K(\designMat, \auginp') \in \R^{n_t \times 1}$. The posterior standard deviation at $\auginp \in \dsnSpace$ is given as $\sigma_{\dataset}(\auginp) := \sqrt{k_{\dataset}(\auginp, \auginp)}$. 
Selecting the hyperparameters $\hyperpara := (\boldsymbol{\ell}, \sigma_f)$ has a strong influence on the GP posterior. In an empirical Bayesian approach the hyperparameters are selected by maximizing the log marginal likelihood. Let $\mathbf{K}_{\hyperpara} := K(\designMat,\designMat \mid \hyperpara) + \sigma_n^2\id_{n_t}$.
Then, the log marginal likelihood is given as
\begin{align}
    \log \pdf{}(\outvec \vert \designMat, \hyperpara) = -\frac{1}{2} (\outvec-\meaneval)^T \mathbf{K}_{\hyperpara}^{-1} (\outvec-\meaneval) - \frac{1}{2} \log \mid \mathbf{K}_{\hyperpara} \vert - \frac{n_t}{2}\log 2\pi. \label{eq:GPMargLike}
\end{align}
Thus, optimal hyperparameters are obtained by 
$
\hyperpara^* = \argmax_{\hyperpara} \log \pdf{}(\outvec \vert \designMat, \hyperpara)
$
. This procedure balances data fit against model complexity and is commonly
associated with the Occam's-razor property of GPs
\citep{rasmussen2000occam}. 

% \subsection{Sequential optimization}
\section{Mono-level surrogate-based robust optimization}
\label{sec:MonoLevel}
% Define the mono-level GP model
% GP surrogate in augmented design space
In the following, we introduce the proposed mono-level robust optimization (MLRO) approach.
The central component of the MLRO approach is a GP surrogate on the augmented design space $\mathcal{Z}$. Conditioned on the available training data $\dataset$, it models the expensive response as 
\begin{align}
    \mdl(\auginp) \approx \surmdl{}(\auginp) \sim \GP(m_{\dataset}(\auginp), k_{\dataset}(\auginp, \auginp')) \qquad
    \auginp,\auginp' \in \augDomain.
\end{align}
% How to estimate the Quantile and its uncertainty using the GP
This GP is used to estimate the output quantile and to quantify the epistemic uncertainty in this estimate induced by limited surrogate training data. To explicitly represent the epistemic uncertainty we introduce the probability space $(\Theta, \mathcal{B}_{\Theta}, \P_{\Theta})$, whose outcomes index
posterior sample functions of the GP. A GP is a random process $\surmdl{}: \augDomain \; \times \; \Theta \rightarrow \R$ and $\surmdl{}(\cdot \vert \theta) := \surmdl{}(\cdot, \theta)$ denotes a realization of this process for $\theta \in \Theta$.
The surrogate predicted random performance measure is then given as
\begin{align}
    \widetilde{\rndouts}_{\inp}(\omega,\theta)
    &:=
    \surmdl{}(
        \inp+\rnddsn(\omega),
        \rndinp(\omega)
        \mid \theta
    ).
    \label{eq:surrogateRandomOutput}
\end{align}
The dependence on $\omega$ represents aleatory uncertainty in the model
inputs, whereas the dependence on $\theta$ represents epistemic
uncertainty in the GP surrogate. 
To retain the epistemic uncertainty, the surrogate-based quantile estimate is treated as a random variable over posterior GP realizations.
For a fixed $\theta \in \Theta$, we thus define
% A realization of $\widetilde{Q}_{\tau}$ is obtained by sampling the GP and computing the corresponding quantile
\begin{align}
    \widetilde{Q}_{\tau}(\inp,\theta) := \cdf{\widetilde{\rndouts}_{\inp}(\cdot, \theta)}^{-1}(\tau).
\end{align}
where $\cdf{\widetilde{\rndouts}_{\inp}(\cdot,\theta)}$ denotes the CDF of the random variable $\widetilde{\rndouts}_{\inp}(\cdot, \theta)$ defined on $(\Omega,\mathcal{B}_{\Omega})$. The mean and the standard deviation of the surrogate-based quantile estimate are
\begin{align}
    \mu_{\widetilde{Q}_{\tau}}(\inp) &= \E_{\mathbb{P}_{\Theta}}[\widetilde{Q}_{\tau}(\inp,\theta)] \label{eq:meanQuant},\\
    \sigma_{\widetilde{Q}_{\tau}}(\inp) &= \V_{\mathbb{P}_{\Theta}}[\widetilde{Q}_{\tau}(\inp,\theta)]^{1/2} \label{eq:stdQuant}.
\end{align}
In practice, (\ref{eq:meanQuant}) and (\ref{eq:stdQuant}) are approximated by a finite sample of $m$ GP realizations. Together with a sampling-based estimation of the quantile the procedure is given as below: 
\begin{enumerate}
    \item Draw $n$ samples
    $\jrndinp(\omega_i) = (\rnddsn(\omega_i),\rndinp(\omega_i))$,
    $i=1,\ldots,n$ and construct the corresponding augmented test points
    \begin{align}
        \auginp_i
        :=
        \left(
            \inp+\rnddsn(\omega_i),
            \rndinp(\omega_i)
        \right),
        \qquad i=1,\ldots,n.
    \end{align}

    \item Draw $m$ realizations of the GP jointly at the test points
    $\{\auginp_i\}_{i=1}^{n}$. For the $j$th realization, this yields
    \begin{align}
        \tilde{\relouts}_i^{(j)}
        :=
        \surmdl{}(\auginp_i \mid \theta_j),
        \qquad
        i=1,\ldots,n
    \end{align}

    \item Sort the outputs of each GP realization, $\tilde{\relouts}_{(1)}^{(j)} \leq \cdots \leq \tilde{\relouts}_{(n)}^{(j)}$ and estimate its $\tau$-quantile by
    \begin{align}
        \tilde{\quant}_{\tau}^{n,(j)}(\inp)
        :=
        \tilde{\relouts}_{(\ceil{n\tau})}^{(j)}.
    \end{align}

    \item Estimate the mean and standard deviation of the
    surrogate-based quantile as
    \begin{align}
        \mu_{\widetilde{Q}_{\tau}}^{n,m}(\inp)
        &:=
        \frac{1}{m}
        \sum_{j=1}^{m}
        \tilde{\quant}_{\tau}^{n,(j)}(\inp),
        \label{eq:sampleEQ}\\
        \sigma_{\widetilde{Q}_{\tau}}^{n,m}(\inp)
        &:=
        \sqrt{
            \frac{1}{m-1}
            \sum_{j=1}^{m}
            \left(
                \tilde{\quant}_{\tau}^{n,(j)}(\inp)
                -
                \mu_{\widetilde{Q}_{\tau}}^{n,m}(\inp)
            \right)^2
        }.
        \label{eq:sampleStdQ}
    \end{align}
\end{enumerate}

% Figure
% Mono-level GP on the left --> Quantile estimate on the right  - Highlight that the GP distribution is not normal anymore
\begin{figure}
	\centering
	\includegraphics[width=\textwidth]{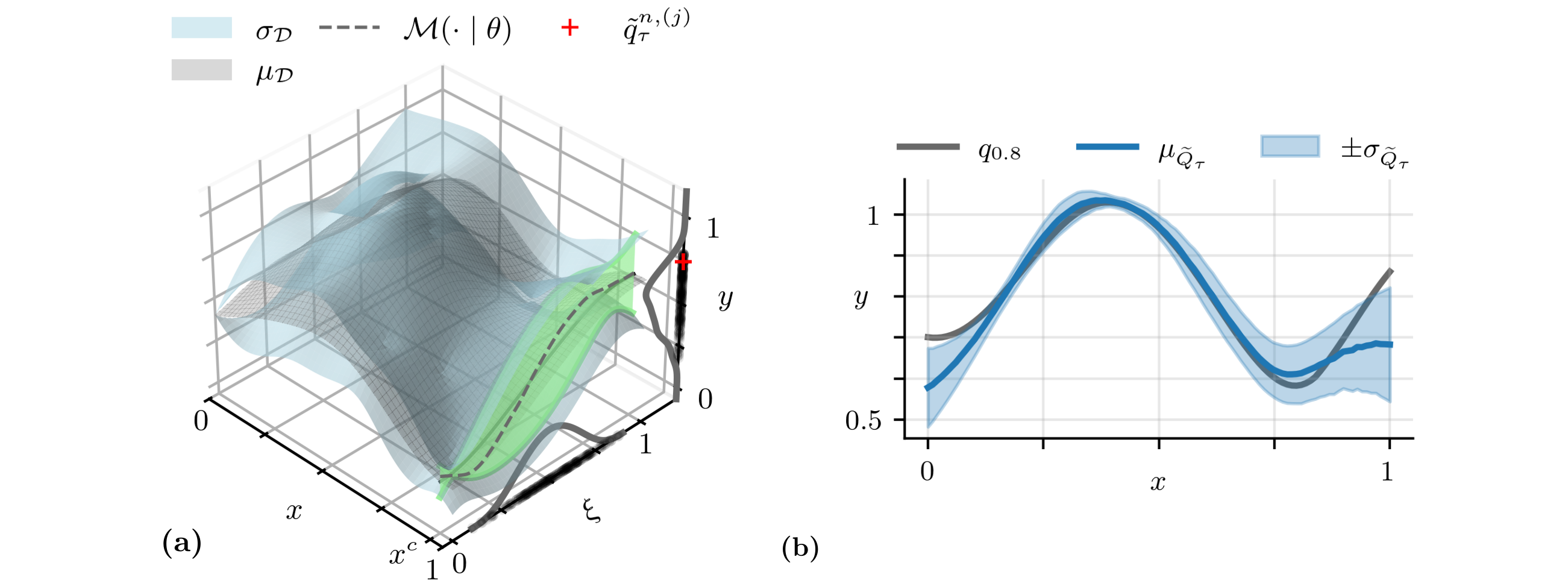}
	\caption{Illustration of the mono-level GP surrogate and quantile estimation at one stage of sequential infill. \textbf{(a)}: GP surrogate in augmented design space for $d_{\inps}=1$ and $d_{\rndinps} = 1$. Additional illustration of the quantile estimation procedure at candidate point $\inp^{c}$. A realization of the GP (dashed line) is evaluated at sample points of the uncertain inputs to obtain an output sample. The empirical quantile of the output sample is denoted with a red cross. \textbf{(b)}: Mean and standard deviation of the surrogate quantile prediction over $\inps$ (blue line and light blue area). The reference quantile obtained from brute force MC simulation is depicted as gray line. }
	\label{fig:MonoApproach}
\end{figure}

The cost of estimating $\mu_{\widetilde{Q}_{\tau}}$ and $\sigma_{\widetilde{Q}_{\tau}}$ is governed by the cost of drawing $m$ samples from the GP at $n$ test points. Using equations \eqref{eq:sampleEQ} and \eqref{eq:sampleStdQ} enables the estimation of the quantile and additionally quantifies the epistemic uncertainty induced by the GP surrogate model. These equations form the basis of the surrogate-based optimization approach proposed in this work.

Figure \ref{fig:MonoApproach}  illustrates the mono-level surrogate and the associated quantile estimation procedure for $d_{\inps}=1$ and $d_{\rndinps} = 1$. The left plot shows sampling-based quantile estimation at a candidate design point $\inps^{c}$. The right plot shows the mean and standard deviation of the surrogate-based quantile estimate, together with a reference quantile obtained by brute-force MC simulation, across $\inps$.

\subsection{Algorithm}
% Steps
% 1. DOE - Initial dataset --> Posterior in augmented design space
The proposed mono-level optimization approach is summarized in Algorithm \ref{alg:MonoLevelRO}. The starting point is a space-filling initial design of experiments in the augmented design space. To this end, a quasi-random low-discrepancy sequence, such as a Sobol' sequence, is used to generate input samples $\auginp^{(i)}$. Evaluating the model at these points yields the initial dataset 
\begin{equation}
    \augDataset := \{(\auginp^{(i)}, \mdl(\auginp^{(i)})) \ \vert \ i = 1, \dots, n_{\text{init}}\}.
\end{equation}
% 2. Identify initial "best q"
Conditioning the GP prior on $\augDataset$ yields a posterior GP that reflects the current belief of the model behavior in the augmented design space. For compact notation in Algorithm \ref{alg:MonoLevelRO}, hyperparameter learning is implicitly included in the posterior inference. The GP posterior mean prediction is used to estimate the minimum quantile $\tilde{\quant}^*$ over the design space. This value serves as initial reference value for the acquisition functions.

% 3. Two stage infill policy
%     - One optimization to determine infill point in the design domain 
%     - One optimization to determine infill points in the stochastic domain
The optimization algorithm proceeds for at most $n_{\text{seq}}$ sequential iterations, adaptively improving the quantile prediction and eventually identifying an optimal design. Each sequential iteration is divided into two stages. In an outer stage, a candidate point $\inp^c$ in the design space is selected. In an inner stage, candidate points $\auginp^c$ are selected in a sub-region of the augmented design space centered at $\inp^c$. At these points the model is evaluated and the GP and the quantile estimate are updated.

% Outer iteration
The goal of the outer stage is to identify the optimal designs based on the quantile estimates \eqref{eq:sampleEQ} and \eqref{eq:sampleStdQ} provided by the surrogate. To this end, we adapt commonly used Bayesian optimization acquisition functions to our setting. An expected-improvement-type criterion balances exploration and exploitation during the early stages of the optimization. In later stages, local acquisition functions restrict the search region around promising designs to accelerate convergence. Further details are provided in Section \ref{ssec:acfn}.

\begin{algorithm}[t!] 
    \caption{Mono-level robust optimization in augmented design space}
    \label{alg:MonoLevelRO}
    \begin{algorithmic}[1]
        \State \textbf{Input:} initial dataset $\augDataset$, $\GP_{\text{prior}}$ 
        \State $\GP_{\text{post}} \gets$ Posterior $(\augDataset, \GP_{\text{prior}})$
        \State $\inp^{*} \gets \argmin_{\inp \in \dsnSpace} \mu_{\widetilde{Q}_{\tau}}(\inp)$ 
        \State $\tilde{q}^*_{\tau} \gets \mu_{\widetilde{Q}_{\tau}}(\inp^{*})$
        \Repeat
            \Comment{Design-candidate selection}
            % \State $\para^{(\inp)}$ \quad \gets \, infer hyperparameters $(\dataset)$
            \State $\inp^c \gets $ Outer acquisition $(\GP_{\text{post}}, \tilde{q}^*_{\tau})$ 
            \State $\tilde{q}^c_{\tau} \gets \mu_{\widetilde{Q}_{\tau}}(\inp^c)$
            \For{$r = 1,\ldots,n_{\mathrm{inner}}$}
                \Comment{Augmented-space infill}
                \State $\auginp^c \gets$ Inner acquisition $(\GP_{\text{post}}, \inp^c, \tilde{q}^c_{\tau})$ 
                \State $\relouts \gets \mdl(\auginp^c)$
                \State $\augDataset \gets $ $\augDataset \cup \{ (\auginp^c, \relouts) \}$
                \State $\GP_{\text{post}} \gets $ Posterior $(\augDataset, \GP_{\text{prior}})$
                \State $\tilde{q}^c_{\tau} \gets \mu_{\widetilde{Q}_{\tau}}(\inp^c)$               
            \EndFor
            \State $\tilde{q}^*_{\tau} \gets \mu_{\widetilde{Q}_{\tau}}(\inp^{*})$ \label{alg:reeval}
            \Comment{Reevaluate optimum}
            \If{$\tilde{q}^c_{\tau}$ < $\tilde{q}^*_{\tau}$}
                \State $\tilde{q}^*_{\tau}, \inp^* \gets $ $\tilde{q}^c_{\tau}, \inp^c$
            \EndIf
        \Until{outer stopping criterion satisfied or budget exhausted $(n_{\mathrm{seq}})$}
        \State \Return{optimum $(\inp^*, \tilde{q}^*_{\tau})$}
    \end{algorithmic}
\end{algorithm}
% Inner iteration
The goal of the inner stage is to improve the surrogate-based quantile estimate at $\inp^c$ and consequently reduce its epistemic uncertainty as indicated by \eqref{eq:sampleStdQ}.  
We therefore use acquisition functions tailored to the uncertainty quantification and in particular to quantile estimation. A popular choice is the maximization of the probability of misclassification \eqref{eq:ProbMiss}. The search space for the inner iteration is the local augmented design space $\augDomain_c = \dsnSpace'_c \, \times \, \mathcal{S}'$, where $\dsnSpace'_c = \prod_{i = 1}^{\din} [\inps^c_i + \rnddsns_i^l, \inps^c_i + \rnddsns_i^u]$. Thus, $\dsnSpace_c'$ contains the realizable perturbed designs associated with the nominal candidate design $\inp^c$, while $\mathcal{S}'$ embeds the uncertain model parameters.
Optimization yields a new candidate point $\auginp^{c} \in \augDomain_c$ at which the model is evaluated. The model evaluation is appended to the dataset and the GP and the quantile prediction are updated. Since the inner iteration involves the computationally expensive model evaluations, it is terminated after only a few iterations based on a fixed iteration limit. In the early stages of the algorithm, the objective is not to obtain a highly accurate quantile estimate at a single candidate design. Instead, the method performs a stepwise reduction of the uncertainty in the quantile estimate, while the outer iteration determines where additional information should be acquired in the design space. In principle, the outer-stage acquisition function may select the same candidate design in multiple sequential iterations, thereby allocating further evaluations to refine its local quantile estimate.

The outer iteration terminates when convergence conditions are met or the computational budget is exhausted. Let $\inp^{*}_{(i)}$ and $\tilde{\quant}_{(i)}^{*}$ denote the current best design
and its estimated quantile after outer iteration $i$, respectively. We monitor the change in the robustness measure, $\left| \tilde{\quant}_{(i)}^{*} - \tilde{\quant}_{(i-1)}^{*} \right| $, as well as the change in the optimal design point, $\| \inp^{*}_{(i)} - \inp^{*}_{(i-1)} \|_2 $. 
After each inner iteration, the quantile estimate at the current best design is reevaluated. This reevaluation is required because the mono-level GP surrogate is global: a new observation affects the posterior and thus potentially affects the quantile estimate at any point in $\dsnSpace$. 

\subsection{Acquisition policies for sequential infill}
\label{ssec:acfn}
% Separate steps: Design space aquisition policy / Stochastic domain aquisition policy
This section discusses the acquisition policies used in the sequential optimization procedure. These policies are a key component of the sequential optimization approach since they decide where the expensive model is evaluated next. 
The goal is to construct tailored policies for specific tasks, e.g., global optimization or uncertainty quantification. The information basis for the policies is a GP posterior that reflects the epistemic uncertainty due to the limited amount of training data. Bayesian decision theory is then applied to guide decision-making under uncertainty.
% Goal: intelligent policies, that minimize the number of model evaluations
\paragraph{Outer-stage acquisition for design optimization.}
In the design space, the objective is to identify candidate designs with low $\tau$-quantiles.
The mono-level surrogate induces a random quantile estimate $\widetilde{Q}_{\tau}$ where the uncertainty is induced by the epistemic uncertainty of the GP surrogate. Let $\tilde{\quant}^*$ denote the current best quantile estimate. The improvement achieved by a candidate design $\inp$ is defined as
\begin{align}
    I(\inp):= \max \{ \tilde{\quant}^*_{\tau} - \widetilde{Q}_{\tau}(\inp), 0 \}. \label{eq:improvement}
\end{align}
Because $\widetilde{Q}_{\tau}$ is a random quantity, the improvement in \eqref{eq:improvement} is also random. The expected improvement (EI) is then given by 
\begin{align}
    \E_{\P_{\Theta}}[I(\inp)] = \E_{\P_{\Theta}} [ \max\{\tilde{\quant}^*_{\tau} - \widetilde{Q}_{\tau}(\inp), 0 \}]. \label{eq:EI}
\end{align}
In general, $\widetilde{Q}_{\tau}$ is non-Gaussian, since it is obtained by propagating the epistemic uncertainty of the GP through the quantile estimator, which is a non-linear operation. 
When the quantile estimate is only mildly non-Gaussian, a Gaussian approximation can be applied, i.e., $\widetilde{Q}_{\tau}(\inp) \overset{\mathrm{approx.}}{\sim} \mathcal{N}(\mu_{\widetilde{Q}_{\tau}}(\inp), \sigma_{\widetilde{Q}_{\tau}}^2(\inp))$. In this case, the classical closed-form EI expression applies
%Expected improvement
\begin{align}
    \E_{\P_{\Theta}}[I(\inp)] \approx a_{\text{EI}}(\inp) = (\tilde{\quant}^* - \mu_{\widetilde{Q}_{\tau}}(\inp)) \Phi \bigg(\frac{\tilde{\quant}^* - \mu_{\widetilde{Q}_{\tau}}(\inp)}{\sigma_{\widetilde{Q}_{\tau}}(\inp)}\bigg) + \sigma_{\widetilde{Q}_{\tau}}(\inp) \phi \bigg(\frac{\tilde{\quant}^* - \mu_{\widetilde{Q}_{\tau}}(\inp)}{\sigma_{\widetilde{Q}_{\tau}}(\inp)}\bigg) \label{eq:AqfEISurr}.
\end{align}
where $\Phi$ and $\phi$ denote the CDF and PDF of the standard normal distribution, respectively. The next candidate design is then obtained by solving a constrained optimization problem
\begin{align}
    \inp^c = \argmax_{\inp \in \dsnSpace} a_{\mathrm{EI}}(x) \quad s.t. \quad \quad \P_{\Omega}[c_j(\inp + \rnddsn) \leq 0] \geq 1 - \eta , \ j = 1, \dots , d_c.
\end{align}
%Both variants are compared in Section AAA. 

% EM
The EI acquisition function balances exploration of poorly sampled regions against exploitation of regions that are expected to improve upon the current optimal design. This is especially important in early stages of the optimization, when the surrogate is based on sparse training data. In the later stages of the optimization the focus can shift towards exploitation.

A purely exploitative policy selects the design with the smallest posterior
mean quantile estimate. This acquisition function is called expected minimum (EM) in the following and defined as
\begin{align}
    a_{\text{EM}}(\inp) = -\mu_{\widetilde{Q}_{\tau}}(\inp). \label{eq:AqfEM}
\end{align}
% TR
In the final stages of the optimization the goal is to refine the solution within an already identified promising region. For this purpose, we use a local trust region (TR) policy. Let
\begin{align}
    \mathcal{T}(\inp^*,\boldsymbol{\rho})
    &:=
    \left\{
        \inp \in \dsnSpace
        \,\middle|\,
        |x_i-x_i^*| \leq \rho_i,
        \quad i=1,\ldots,\din
    \right\}
\end{align}
denote a box-shaped trust region centered at the current optimal design
$\inp^*$, with radius vector
$\boldsymbol{\rho}\in\R_+^{\din}$. The next candidate point is selected by
\begin{align}
    \inp^c
    =
    \argmin_{\inp \in \mathcal{T}(\inp^*,\boldsymbol{\rho})}
    \mu_{\widetilde{Q}_{\tau}}(\inp) \quad s.t. \quad \quad \P_{\Omega}[c_j(\inp + \rnddsn) \leq 0] \geq 1 - \eta , \ j = 1, \dots , d_c.
    \label{eq:AqfTrust}
\end{align}
If a new optimal design is identified, the center of the trust region is
updated accordingly. The trust-region radius can subsequently be expanded
or contracted according to a prescribed update rule. We refer to
\citep{alexandrov1998trust} for further details on trust-region methods.

In the actual implementation the finite sample estimators \eqref{eq:sampleEQ} and \eqref{eq:sampleStdQ} of $\mu_{\widetilde{Q}_{\tau}}$ and $\sigma_{\widetilde{Q}_{\tau}}$ are used. During EI and EM iterations we adopt a coarse approximation using $n=m=5000$. In the final TR stage the sample size is increased to $n = 1 \times 10^5$ for accurate quantile estimates.
% Stoachstic infill citeria
% - SE
% - Ufun --> Maybe discuss extension of the classical U function
\paragraph{Inner-stage acquisition for uncertainty quantification.}
For each candidate design $\inp^c$, the inner iteration aims to improve the surrogate-based estimation of the robustness measure. To this end, the GP surrogate must be accurate within $\augDomain_c$, that is relevant to uncertainty propagation at $\inp^c$. The accuracy of the surrogate in $\augDomain_c$ directly affects the accuracy of the resulting robustness measure estimate. Different acquisition functions have been proposed to adaptively refine a GP surrogate used for uncertainty propagation. 

A simple uncertainty-reduction criterion selects the point of largest GP posterior standard deviation within $\augDomain_c$. This criterion is particularly suitable for robustness measures based on global response statistics, e.g., a weighted combination of the mean and standard deviation. The corresponding maximum-variance infill point is defined by   
\begin{align}
    \auginp^c = \argmax_{\auginp \in \augDomain_c} \sigma_{\augDataset}(\auginp). \label{eq:AfnStochInfillSE}
\end{align}

Next, we discuss an acquisition function specifically designed to improve quantile estimation. A $\tau$-quantile can be seen as a threshold point that separates an output sample into specific proportions. When a surrogate model is used for quantile estimation, it must accurately classify whether input samples are below or above the estimated quantile threshold. 

For a fixed estimated quantile threshold at the candidate design
$\inp^c$, the posterior probability of misclassification at an augmented
input $\auginp \in \augDomain_c$ is
\begin{align}
    P_{\mathrm{M}}(\auginp) := \Phi\!\left(-\frac{\left|\tilde{q}_{\tau}^{\,c}-\mu_{\augDataset}(\auginp)\right|}{\sigma_{\augDataset}(\auginp)}\right),
    \label{eq:ProbMiss}
\end{align}
where $\tilde{q}_{\tau}^{\,c}$ denotes the current surrogate-based estimate of the $\tau$-quantile at $\inp^c$, and $\Phi$ is the CDF of the standard normal distribution.
Maximizing the probability of misclassification is equivalent to minimizing the so called U-function \citep{echard2011ak} 
\begin{align}
    U(\auginp) :=\frac{\left|\tilde{q}^{c}-\mu_{\augDataset}(\auginp)\right|}{\sigma_{\augDataset}(\auginp)}.
\end{align}
Accordingly, the inner-stage infill point is selected as
\begin{align}
    \auginp^c :=\argmin_{\auginp \in \augDomain_c} U(\auginp;\inp^c).\label{eq:AfnStochInfillUFun}
\end{align}
This criterion targets regions where the posterior GP mean is close to the estimated quantile threshold and the posterior uncertainty is large. These regions are prone to misclassify input samples and important to refine for accurate quantile estimation.  

\paragraph{Heuristics for design of experiment.}
\label{ssec:Heuristics}
In theory, the infill iterations discussed above may be continued until convergence is observed. In practice, expensive computational models are typically subject to a fixed computational budget. In this case, heuristic budget allocation rules are required to distribute available model evaluations across the initial design and the different infill stages. 

For the initial design of experiments a rule of thumb is to use between $5d$ and $10d$ samples, where $d$ is the dimensionality of the problem \citep{jones1998efficient, loeppky2009choosing}. In this work we adopt a standard choice and set $n_{\mathrm{init}} = 10 (\din + \stochin)$ unless stated otherwise. For the sequential infill iterations, we use $n_{\mathrm{seq}} = 2 n_{\mathrm{init}}$. Of these iterations, \qty{60}{\percent} use EI-based infill, \qty{20}{\percent} EM-based infill and \qty{20}{\percent} TR-based infill \citep{sabater2022development}.

For the uncertainty quantification infill, we combine the maximum-variance criterion in \eqref{eq:AfnStochInfillSE} with the U-function criterion in \eqref{eq:AfnStochInfillUFun}. Within each inner stage, the first $n_{\mathrm{V}}$ infill points are selected using the maximum-variance criterion. The subsequent $n_{\mathrm{U}}$ points are selected using the U-function criterion. The computational budget is limited and set to $n_{\mathrm{V}} = \bigl \lceil \stochin / 3 \bigr \rceil$ and $n_{\mathrm{U}} = \bigl \lceil \stochin / 2 \bigr \rceil$ defining the number of inner stage model evaluations $n_{\mathrm{inner}} = n_{\mathrm{V}} + n_{\mathrm{U}}$. This limited budget exploits a strength of the mono-level approach: the inner stage does not need to converge at a single candidate design. Instead, the outer stage can return to the same design point in a later iteration and allocate additional evaluations to refine its quantile estimate, if necessary. The overall number of model evaluations of the MLRO approach is thus given by $n_t = n_\mathrm{init} + n_\mathrm{seq}n_{\mathrm{inner}}$

\subsection{Efficient GP posterior sampling}
\label{ssec:effSampling}

The computational cost of the mono-level robust optimization approach is dominated by model evaluations of $\mdl$, which may require solving complex simulation codes. However, optimizing the design space acquisition function can also be computationally demanding. Its evaluation requires estimating the quantile and the uncertainty of this estimate at many points in the design space. In the mono-level approach, the estimators in \eqref{eq:sampleEQ} and in \eqref{eq:sampleStdQ}, require $m$ GP posterior realizations evaluated at $n$ test points. 

Classical methods for sampling from a GP posterior scale poorly with the number of test points, as they require a Cholesky decomposition of the posterior covariance matrix with computational complexity of $\mathcal{O}(n^3)$. This leads to the conflicting goals: On the one hand, large sample sizes $n$ and $m$ improve the accuracy of the quantile estimate. On the other hand, large sample sizes drastically increase the computational cost of optimizing the acquisition function and limit the scalability of the mono-level approach.

To address this challenge, we adopt the efficient posterior sampling method introduced in \cite{wilson2020efficiently}. The method combines pathwise sampling with a finite-dimensional weight-space representation of a stationary GP prior. It enables sampling with computational cost that scales linearly in the number of test points, at the cost of incurring a small kernel approximation error. The key components of the approach are summarized below:

% Pathwise sampling
Generalizing Matheron’s rule \cite{chiles2012geostatistics, doucet2010note} from Gaussian random variables to GPs gives the following representation of the posterior: 
\begin{align}
    f_{\dataset}(\cdot) \overset{d}{=}
    \underbrace{f(\cdot)}_{\text{prior}}
    +
    \underbrace{k(\cdot, \designMat) \left(K(\designMat,\designMat) + \sigma^2_n \id_{n_t}\right)^{-1} \left(\outvec - \mathbf{f}_{\designMat} - \boldsymbol{\varepsilon}\right)}_{\text{update}}.
    \label{eq:pathwiseUpdate}
\end{align}
Here $f$ is a prior GP sample and $\mathbf{f}_{\designMat} = [f(\auginp^{(1)}),\ldots,f(\auginp^{(n_{\mathrm{t}})})]^\top$ is the prior evaluated at the training inputs. This expression decomposes a posterior sample into a prior sample and a data-dependent correction term. The correction accounts for the residual between the observations $\outvec$ and $\mathbf{f}_{\designMat}$. Thus, samples from the posterior can be generated by drawing prior samples and subsequently applying the update in \eqref{eq:pathwiseUpdate}. This avoids constructing the posterior covariance matrix or its Cholesky decomposition. However, it still requires drawing samples from the prior distribution.

% Weight space view
To efficiently draw samples from the prior distribution a weight-space representation of a GP is adopted. A positive-definite kernel $k$ induces an inner product in a reproducing kernel Hilbert space $\mathcal{H}$ with a corresponding feature map $\varphi: \augDomain \rightarrow \mathcal{H}$. In the weight-space view, this kernel is approximated by replacing the exact, infinite-dimensional feature representation with a finite-dimensional basis expansion $\psi: \augDomain \rightarrow \R^r$ and thus 
\begin{align}
    k(\auginp, \auginp') = \langle \varphi(\auginp), \varphi(\auginp') \rangle_{\mathcal{H}} \approx \psi(\auginp)^T \psi(\auginp').
\end{align}
As a result, the corresponding GP prior is approximated by a Bayesian linear model in this finite feature space. The GP in weight-space representation thus reads 
\begin{align}
f(\cdot) \approx \sum_{i = 1}^r w_i \psi_i(\cdot),
\end{align}
with $\psi_i(\cdot)$ denoting the basis functions and $w_i \sim \mathcal{N}(0,\sigma_f^2)$ independent random weights. 
%The basis functions are selected to match the kernel of the prior distribution. 
For stationary covariance functions, Bochner’s theorem \cite{rahimi2007random} shows that an appropriate $r$-dimensional basis can be constructed using random Fourier features
\begin{align}
\psi_i(\auginp) = \sqrt{2/r} \cos(\bs{\kappa}_i^T \auginp + \beta_i). 
\end{align}
Here, the phase is sampled according to $\beta_i \sim U(0,2\pi)$ and the frequency vector $\bs{\kappa}_i$ is sampled from the spectral density associated with the kernel. This spectral density is obtained by computing the Fourier transform of $k(\auginp - \auginp')$, i.e. 
\begin{align}
    p(\bs{\kappa}) = \frac{1}{(2\pi)^{d_z}} \int_{\R^{d_z}} e^{- j\bs{\kappa}^T \Delta} k(\Delta) d\Delta \quad \quad \text{with} \quad \Delta = \auginp - \auginp'.  
\end{align}
For an isotropic Mat\'ern kernel with length scale $l$, the corresponding
spectral density is, up to a normalizing constant, \citep[p.84]{williams2006gaussian}
\begin{align}
    p(\bs{\kappa}) \propto \big(\frac{2\nu}{l^2} + \| \bs{\kappa} \| ^2 \big)^{-(\nu + d_z/2)}. \label{eq:SpecDens}
\end{align}
For the anisotropic kernel used in this work, the frequency vectors are scaled componentwise by the corresponding inverse length scales. Here we see that the hyperparameters of the kernel influence its spectral density and thus the frequencies of the random Fourier features. For practical implementation, the spectral density (\ref{eq:SpecDens}) can be rewritten as multivariate Student's t-distribution, which can be sampled via a Gaussian-Gamma mixture approach. 

Under this weight-space representation, sampling from a GP prior is very efficient since it only requires evaluating the basis functions $\psi_i$ at the sample locations. In fact, sampling now scales linearly with $n$, at the cost of introducing a small error due to the finite-dimensional basis approximation ($r = 1024$) of the kernel. After efficient sampling of the prior, posterior samples are obtained by conditioning on the data using Matheron's rule as stated in \eqref{eq:pathwiseUpdate}. This conditioning step is cheap since we usually deal with a small training dataset of size $n_{t}$ and computing Cholesky decompositions with complexity $\mathcal{O}(n_{t}^3)$ is feasible. 

\subsection{Illustrative 1D example}
\begin{figure}
	\centering
	\includegraphics[width=300pt]{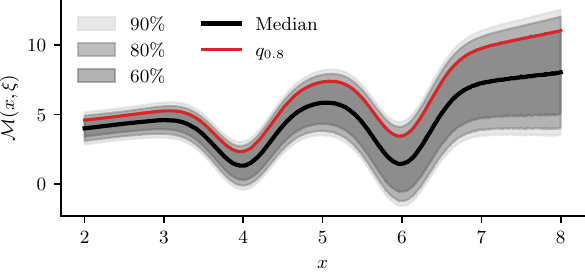}
	\caption{Analytical benchmark model used as illustrative example. The \qty{60}{\percent}, \qty{80}{\percent}, and \qty{90}{\percent} central probability regions are shown as gray shaded areas. The \qty{80}{\percent} quantile is depicted as red line.} 
	\label{fig:Sabater1d}
\end{figure}

Finally, we illustrate the proposed MLRO approach using an analytical benchmark model.
%Introduce the equation
The benchmark model adopted from \citep{sabater2021bayesian}, is given by
\begin{align}
    \relouts = \mdl(\inps, \rndinp) = 3 - 4 \exp(-4(\inps-4)^2)-5.2\exp(-4(\inps-6)^2)+\frac{\inps-a}{b-a}\rndinps_1 + \frac{b-\inps}{b-a}\rndinps_2\label{eq:1D_test}.
\end{align}
The model has a one-dimensional design space with $\inps \in [a, b]$ and $a=2.0$ and $b=8.0$ and two uncertain inputs $\rndinps_1 \sim U(0,10)$ and $\rndinps_2 \sim \mathcal{N}(1.01,0.71^2)$. This example does not consider uncertainty in the design variables.
% Figure of the stochastic model output
Figure \ref{fig:Sabater1d} shows the median and the associated \qty{60}{\percent}, \qty{80}{\percent}, and \qty{90}{\percent} central probability regions across $\inps$. The \qty{80}{\percent} quantile, $\quant_{0.8}$, depicted as a red line serves as the robustness measure. It has a global minimum near $x \approx 4$ and local minima at $x \approx 6$ and $x = 2$.

\begin{figure}
	\centering
	\includegraphics[width=300pt]{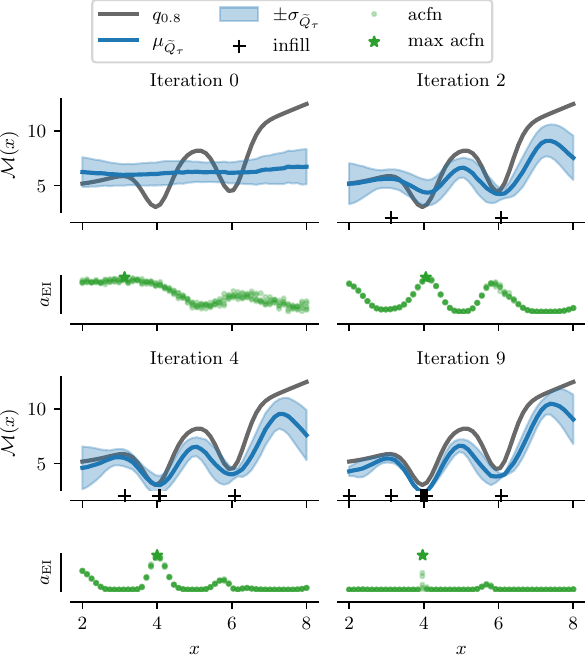}
	\caption{Snapshots of the MLRO approach for different iterations. For each iteration the surrogate quantile estimation (blue) is compared to the brute-force MC quantile estimate (gray). Evaluations of the expected improvement type acquisition function are shown below each quantile prediction; its maximum indicated as green star.}
	\label{fig:evo_sabat1d}
\end{figure}

% Evolution plot
Figure \ref{fig:evo_sabat1d} shows the evolution of the mono-level approach at selected iterations. The estimated quantile is compared with the reference quantile, and the expected-improvement acquisition function is shown, which is used to adaptively refine the surrogate. The MLRO approach is initialized using a dataset $\augDataset$ with $n_{\text{init}} = 10$ samples in the augmented design space.

At iteration $0$, the mono-level surrogate provides only a rough quantile estimate and exhibits substantial epistemic uncertainty. Maximizing the acquisition function yields the infill points in the design space, indicated by black crosses. At each candidate design, two infill points are selected using \eqref{eq:AfnStochInfillSE} and \eqref{eq:AfnStochInfillUFun}. These evaluations refine the surrogate and reduce the uncertainty in the quantile prediction as can be seen in iteration $2$. The algorithm is able to escape local minima by leveraging the exploration capacities of the EI acquisition function. This way the global optimum can be discovered as seen in iteration $4$. Once discovered, the algorithm exploits the global optimum, indicated by a concentrated acquisition function, and adds additional samples in this region (see iteration $9$).  

\begin{figure}
	\centering
	\includegraphics[width=330pt]{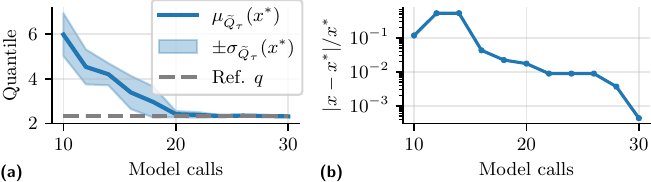}
	\caption{\textbf{(a)}: Evolution of the mean and standard deviation of the surrogate-based quantile estimate over the model evaluations. \textbf{(b)}: Convergence of the estimated optimal design point, measured by the relative error over model evaluations.}
	\label{fig:conv_sabat1d}
\end{figure}

% Convergence plot
Finally, Figure \ref{fig:conv_sabat1d} shows the evolution of the estimated optimal design and its corresponding quantile. The uncertainty in the quantile estimation decreases throughout the optimization, particularly after the global optimum has been identified. To assess convergence of the optimal design point $\inp^*$ we additionally report the relative error over the model evaluations. The results show that the algorithm is able to locate the global optimum, with the relative error decreasing in later iterations.

\newpage

\section{Applications}
\label{sec:App}
In this section, we apply the MLRO approach to analytical test functions and a structural optimization problem. Unless stated otherwise, the maximum number of model evaluations is set according to the budget-allocation heuristics defined in Section~\ref{ssec:Heuristics}.

\subsection{Analytical test functions}
\label{ssec:analtest}
Analytical test functions serve as common benchmarks to assess the performance of optimization algorithms. Here, we consider common test functions for optimization problems and augment their inputs with stochastic variables. Table \ref{AnalTestFun} lists the selected test functions and summarizes important properties. All test functions considered here are smooth and differ primarily in their dimensionality and optimization landscape.  
\begin{table}[width=.45\linewidth,cols=4,pos=h]
\caption{Analytical test functions for robust optimization.}\label{AnalTestFun}
\begin{tabular*}{\tblwidth}{@{} LCCCL@{} }
\toprule
\textbf{Benchmark} & $\din$ & $\stochin$ & \#Loc. min. & Ref.\\
\midrule
Sabater & 2 & 2 & 8 & \citep{sabater2021bayesian} \\
Branin-Hoo+UQ & 2 & 3 & 2 &  \citep[p.~196]{forrester2008engineering} \\
Hartmann-6+UQ & 6 & 3 & 5 & \citep{hartman1973some} \\
\bottomrule
\end{tabular*}
\end{table}

The Sabater test function \citep{sabater2021bayesian} has a low-dimensional smooth, multi-modal and separable design space. The main challenge associated with this test function are the multiple local minima. In this work, we study the effect of different stochastic model behavior on the MLRO approach. 
%We additionally investigate the effect of localized features in the design domain by varying the width of the valley associated with the global optimum. 

The Branin-Hoo test function \citep[p.~196]{forrester2008engineering} is a smooth, non-convex, anisotropic test function with a narrow curved valley and strong variable coupling, making it a challenging benchmark for optimization algorithms. In the modified version considered here, it has one unique global minimum and the main challenge is to identify the global optimum along the flat valley. In this work we augment it with stochastic parameters to adapt it to the robust optimization setup. 

Finally, we consider the Hartmann-6 function \citep{hartman1973some} a nonlinear, highly nonconvex, and multimodal test function. We augmented this six-dimensional function with three stochastic parameters, resulting in a 9-dimensional problem for the mono-level surrogate. For all test functions we consider the robust optimization problem as defined in \eqref{eq:robOptProb} with $95 \%$ quantile as the robustness measure and no additional inequality constraints.

For each test function, we monitor the relative errors of the current best design variables $\inp_{(i)}^*$ 
\begin{equation}
    \epsilon_{\inp,\mathrm{rel}}^{(i)} = \frac{ \| \inp_{(i)}^* - \inp^* \|_2}{\| \inp^* \|_2}
\end{equation}
and the current best quantile estimate $\tilde{q}_{(i)}^*$
\begin{equation}
    \epsilon_{q,\mathrm{rel}}^{(i)} = \frac{ | \tilde{q}_{(i)}^* - q^* |}{| q^* |}.
\end{equation}
The reference optima $(\inp^*, q^*)$ are computed using a brute-force procedure that combines direct MC quantile estimation with a global genetic optimizer. The MLRO approach is repeated ten times with different seeds to capture different random effects, including scrambled DOE, hyperparameter tuning and sampling-based quantile estimation. We report the individual runs, together with their median and central \qty{80}{\percent} probability interval.

\paragraph{Sabater test function.} The Sabater test function is given as 
\begin{align}
    \mdl(\inp, \rndinp) =& \sum_{i = 1}^2 \Big[ 3 - 4 \exp(-4(\inps_i-4)^2)-5.2\exp(-4(\inps_i-6)^2)+ g(\inps_i, \rndinp) \Big]
    \label{eq:Sabat2d}
\end{align}
with $\inp \in [l, u] \times [l, u]$ and $l=2$, $u=8$, $\rndinps_1 \sim U(0,10)$ and $\rndinps_2 \sim \mathcal{N}(1.01,0.71^2)$. The term $g$ models the stochastic part for which we consider the three cases:
\begin{align}
    g_{\mathrm{1}}(\inps_i, \rndinp) &= \frac{\inps_i-l}{u-l}\xi_1 + \frac{u-\inps_i}{u-l}\rndinps_2, \\
    g_{\mathrm{2}}(\inps_i, \rndinp) &= 0.1\frac{\inps_i-l}{u-l}\rndinps_1^2 + \frac{u-\inps_i}{u-l}\xi_2^2 \quad \text{and} \\
    g_{\mathrm{3}}(\inps_i, \rndinp) &= 0.1\,\frac{\inps_i-l}{u-l}\left(\frac{\rndinps_1-5.0}{1.4}\right)^4\;+\;0.3\,\frac{u-\inps_i}{u-l}\,\rndinps_2^2\;+\;5\,\sin^2\!\left(0.2\,\rndinps_1\,\rndinps_2\right).
\end{align}
The function $g_1$ defines a linear behavior in the stochastic parameters as originally proposed by \citep{sabater2021bayesian}. 
\begin{figure}
	\centering
	\includegraphics[width=300pt]{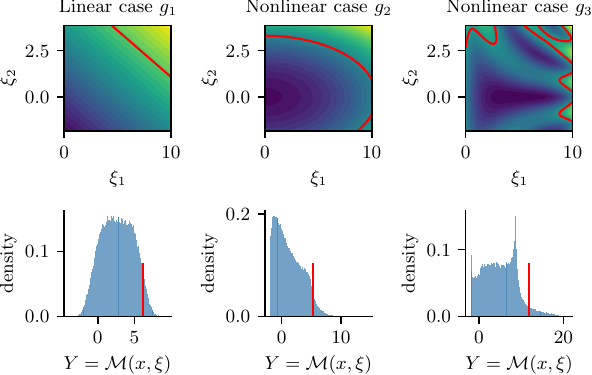}
	\caption{Different stochastic model behaviour of the Sabater test function. The top row depicts the model response over the stochastic parameters of the test function at a nominal design point $\inp = [4.0, 4.0]$. The bottom row depicts the push forward distribution of the input uncertainty to the model output. The red line denotes the $95 \%$ quantile.}
	\label{fig:Sab2dUQDomain}
\end{figure}
The function $g_2$ models a mildly nonlinear behavior by introducing quadratic terms. Finally, $g_3$ represents strongly nonlinear stochastic parameters including interaction effects of $\rndinps_1$ and $\rndinps_2$. The resulting model responses over the stochastic parameters are illustrated in Figure \ref{fig:Sab2dUQDomain} at the design point $\inp = [4.0, 4.0]$, which is close to the optimum. The red contour denotes the set of stochastic inputs for which the model response equals the $95\%$ quantile threshold. Accurate quantile prediction requires the mono-level surrogate to classify whether the corresponding model responses lie below or above. The complexity of this contour increases from $g_1$ to $g_3$. The second row shows the corresponding push-forward distributions of the uncertain inputs. Due to the nonlinearities we observe mildly and strongly skewed distributions. 

\begin{figure}
	\centering
	\includegraphics[width=300pt]{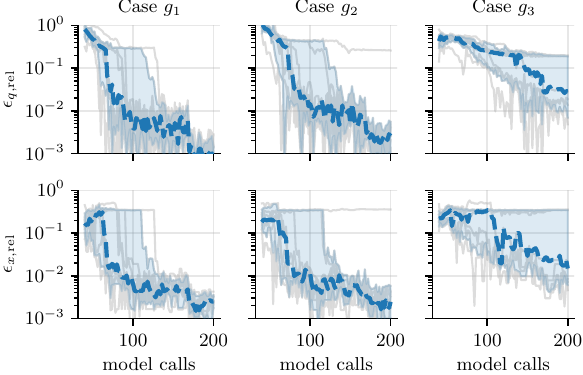}
	\caption{Evolution of MLRO approach for different cases of the Sabater test functions. The top row depicts the evolution of the relative error of the best quantile estimate and the bottom row shows the relative error of the best design point. Grey lines represent individual runs, blue dashed line the median and light blue area the \qty{80}{\percent} central probability region.} 
	\label{fig:Sab2dResults}
\end{figure}

Figure \ref{fig:Sab2dResults} shows the evolution of the MLRO approach for the three Sabater test function variants. Individual runs are plotted in light gray, while the median and central \qty{80}{\percent} probability are displayed in blue and light blue, respectively. The top row shows the relative error in the estimated robustness measure, i.e. the \qty{95}{\percent} quantile. For $g_1$ and $g_2$ the relative error decreases rapidly from \qty{100}{\percent} to below \qty{1}{\percent} within \qty{100} model evaluations. The error then stagnates for a few iterations and is finally further reduced during the TR infill iterations to around \qty{0.2}{\percent}. For the strongly nonlinear model $g_3$, five out of 10 simulation runs get stuck in a local minimum and thus the median is shifted towards higher errors. This observation underlines that the chosen maximal computational budget is not sufficient to solve the case $g_3$ robustly. Adding, e.g., additional initial DOE samples could improve the results as reported in Appendix \ref{app:DOEstudySabat_g3}. 
The second row shows the corresponding relative error in the best design vector. While for the cases $g_1$ and $g_2$ a fast decay can be observed, we see a slower decay for the case $g_3$ for those simulations that identified the global optimum. A detailed evolution plot for a single optimization run for the case $g_2$, as well as the reference and predicted quantile over the design space are provided in Appendix \ref{sec:AddPlots}.        

\begin{figure}
	\centering
	\includegraphics[width=230pt]{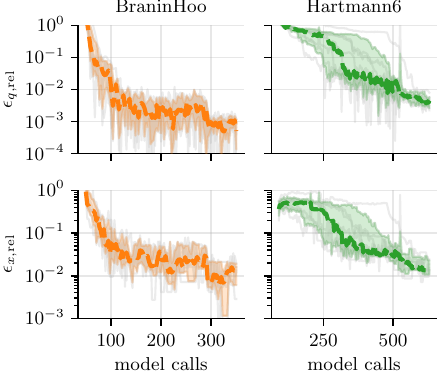}
	\caption{Evolution of MLRO approach for the Branin-Hoo and Hartmann-6 test function. The top row depicts the evolution of the relative error of the best quantile estimate and the bottom row shows the relative error of the best design point. Grey solid lines represent individual runs, dashed line the median and shaded area the \qty{80}{\percent} central probability region.}
	\label{fig:BraninHartmannConv}
\end{figure}

\paragraph{Branin-Hoo test function.} The modified Branin-Hoo test function is given as 
\begin{align}
    \mdl(\inp, \rndinp) =& a (\inps_2 - b\inps_1^2 + c \inps_1 -r)^2 + s(1-t)\cos(\inps_1) + s + 2\inps_1 + \nonumber \\
    &5 \bigg(\frac{\inps_1 - l_1}{u_1 - l_1} \bigg) \sin(\rndinps_1) + 5\bigg(\frac{\inps_2 - l_2}{u_2 - l_2} \bigg) \rndinps_2^2 + 0.1(\inps_1 - 4)^2 \rndinps_3,
    \label{eq:Branin}
\end{align}
with $\inp \in [l_1, u_1] \times [l_2, u_2]$ and $l_1=-5.0$, $u_1=10$, $l_2=0.0$, $u_2=15.0$. The random parameters are uniformly distributed, i.e., $\rndinps_1, \rndinps_2, \rndinps_3 \sim U(-\pi, \pi)$ and the constant parameters of the Branin-Hoo function are $a=1$, $b=5.1/(2\pi)^2$, $c=5/\pi$, $r=6$, $s=10$ and $t=1/(8\pi)$. The left panel of Figure \ref{fig:BraninHartmannConv} shows the evolution of the MLRO approach for the Branin-Hoo test function. The relative error of the quantile estimate falls below \qty{1}{\percent} after approximately \qty{100} model evaluations, whereas the relative error in the design vector decreases more slowly. This slow convergence can be attributed to the narrow curved valley around the global optimum of the Branin-Hoo test function. A visual representation of the reference optimization landscape and the predicted landscape by the MLRO approach is provided in Appendix \ref{sec:AddPlots}. Here, we see that the MLRO approach quickly discovers the region of the global optimum and then slowly moves along the flat valley towards the global optimum. 

\paragraph{Hartmann-6 test function.} The Hartmann-6 test function with augmented stochastic parameters is given as
\begin{align}
    \mdl(\inp, \rndinp) =& -\sum_{i=1}^4\alpha_i \exp \bigg( - \sum_{j=1}^6 A_{ij}(\inps_j - P_{ij})^2 \bigg) + \nonumber \\
    &5 \inps_1 \sin(\rndinps_1) + (\inps_2 -0.5)^2 \rndinps_2^2 + 0.5(\inps_1 - 0.8)^2 \rndinps_3,
\end{align}
with $\inp \in [0,1]^6$ and $\rndinps_1, \rndinps_2, \rndinps_3 \sim U(-\pi, \pi)$. The constants of the test function are defined as 
\begin{align}
    \alpha = &\begin{bmatrix} 1.0 & 1.2 & 3.0 & 3.2 \end{bmatrix}^T \nonumber \\
    A = &\begin{bmatrix} 
10 &  3   &   17   &  3.5   & 1.7 & 8 \\ 
0.05  & 10   &   17   &  0.1  &  8 & 14\\
3 &  3.5  & 1.7   & 10    &  17 & 8\\
17      &  8  & 0.05 & 10  &  0.1  & 14  \\ 
\end{bmatrix} \nonumber \\
P = 10^{-4} &\begin{bmatrix} 
        1312 & 1696&5569& 124& 8283& 5886 \\
        2329 & 4135& 8307& 3736& 1004& 9991 \\
        2348 & 1451& 3522& 2883& 3047& 6650 \\
        4047 & 8828& 8732& 5743& 1091& 381 \\
\end{bmatrix} \nonumber.
\end{align}
The right panel of Figure \ref{fig:BraninHartmannConv} shows the evolution of the MLRO approach for the Hartmann-6 test function. For this higher-dimensional problem the MLRO approach requires around \qty{500} model evaluations to reduce the relative error of the quantile estimate below \qty{1}{\percent}. A detailed evolution plot of the quantile estimate and the current best design vector for a single simulation run is provided in Appendix \ref{sec:AddPlots}. 

\subsection{Bracket model}
% Intro
Finally, we apply the MLRO approach to a structural optimization problem. The goal is to optimize the shape of a bracket structure subject to uncertain load conditions. This optimization problem has been introduced in \citet{archbold2024variational} as a test-case for robust optimization and has been used as a benchmark in \citet{essou2024}. In contrast to the previously investigated analytical test functions, this optimization problem includes constraints as well as uncertainties in the design variables.

\begin{figure}
	\centering
	\includegraphics[width=\textwidth]{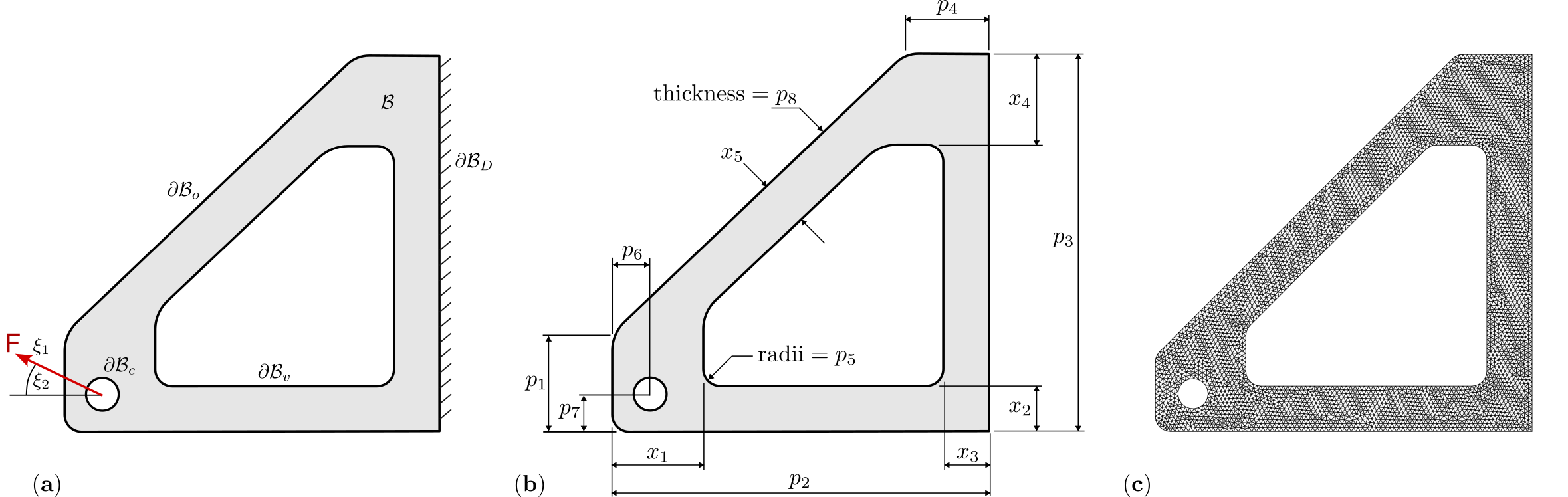}
	\caption{Bracket model setup: \textbf{(a)} Domain, boundaries and loading \textbf{(b)} geometry with design and fixed parameters \textbf{(c)} triangular finite element mesh.}
	\label{fig:BracketModel}
\end{figure}

Figure \ref{fig:BracketModel} depicts the geometry of the bracket which is defined by a design vector $\inp = (\inps_1, \dots, \inps_{5})^\top$ and a vector of fixed parameters $\bs{p} = (p_1, \dots, p_{8})^\top$. Let $\boldsymbol{s} \in \R^2$ denote the spatial coordinate and let $\mathcal{B}\subset\R^2$ denote the bracket domain and $\partial \mathcal{B}$ its boundaries. Along $\partial \mathcal{B}_D$ a homogeneous Dirichlet boundary condition is applied to model a clamped edge. The boundaries $\partial \mathcal{B}_o$ and $\partial \mathcal{B}_v$ are traction-free and therefore satisfy a homogeneous Neumann condition. On the circular boundary $\partial \mathcal{B}_c$, the surface traction $\boldsymbol{g}(\boldsymbol{s}) \in \R^2$ induces a point force loading $\int_{\partial \mathcal{B}_c} \boldsymbol{g} (\boldsymbol{s}) d \boldsymbol{s} = [\xi_{1} \cos(\xi_{2}), \xi_{1} \sin(\xi_{2})]^T$ and a body force $\boldsymbol{b} = [0, \qty{-7.7e-5}{\newton\per\cubic\milli\metre}]$ acts as a constant loading over the domain. As material we consider steel with isotropic material behavior and Young's modulus $E = \qty{2.1e5}{\newton\per\milli\metre\squared}$ and Poisson's ratio $\nu = 0.3$.  

% Design and uncertain variables
The bounds for the design vector $\inp$ and the fixed values of the parameter vector $\bs{p}$ are given in Table \ref{tab:BracketBounds}. The final design is subject to manufacturing imperfections, modeled by the zero-mean perturbation $\bs{\delta} \sim \mathcal{N}(\boldsymbol{0}, \sigma_{\delta}^{2}\id_{\din})$ with standard deviation $\sigma_{\delta} = \qty{0.1}{\milli\metre}$. Moreover the load varies during operation and is subject to uncertainties. In particular, the point force magnitude $\xi_1 \sim \mathcal{N}(5000, 1000^2)$ and the angle of the point force $\xi_2  \sim \mathcal{N}(0, (\pi/6)^2)$ are modeled as random variables.
\begin{table}[h]
\caption{Bounds for the design variables $\inp$ and nominal values for the fixed parameters $\bs{p}$ in \unit{\milli\meter}.}
\label{tab:BracketBounds}
\centering
\begin{tabular}{@{}lccccclcccccccc@{}}
\toprule
\textbf{Design} & $x_1$ & $x_2$ & $x_3$ & $x_4$ & $x_5$
& \textbf{Fixed} & $p_1$ & $p_2$ & $p_3$ & $p_4$ & $p_5$ & $p_6$ & $p_7$ & $p_8$ \\
\cmidrule(lr){2-6} \cmidrule(lr){8-15}
& [60, 100] & [10, 50] & [10, 50] & [60, 100] & [10, 30]
& & 60 & 250 & 250 & 50 & 10 & 25 & 25 & 5 \\
\bottomrule
\end{tabular}
\end{table}

% PDE, target quantity and constraints
The objective for optimization is to minimize the compliance of the structure. Assuming small deformations, the bracket structure is modeled using linear elasticity. Furthermore, the bracket is modeled as a thin plate based on a plane stress assumption. Let $\bs{u}(\boldsymbol{s} ; \inp + \bs{\delta}, \rndinp) \in \R^2$ denote the displacement field indicating that it depends on the (random) design and load vector. The corresponding strain and stress tensors are denoted by $\bs{\varepsilon}(\bs{u}) \in \R^{2\times2}$ and $\bs{\sigma}_s(\bs{u}) \in \R^{2\times2}$. In this study we seek a design $\inp^*$ that minimizes the compliance of the structure
\begin{align}
    \rndouts_{\inp} := \int_{\mathcal B} \bs{\sigma}_s \!\left( \bs{u}(\boldsymbol{s}; \inp+\rnddsn,\rndinp) \right) : \bs{\varepsilon}\!\left(\bs{u}(\boldsymbol{s};\inp+\rnddsn, \rndinp)\right)\,\mathrm d\boldsymbol{s}
\end{align}
subject to a volume constraint
\begin{align}
    c(\inp+\rnddsn) = \int_{\mathcal{B}} d\bs{s} - V_0
\end{align}
with $V_0 = \qty{1.25e5}{\milli\meter\cubed}$. Due to the uncertainty in design and load parameters the compliance is a random quantity itself. The robust objective is defined as the \qty{95}{\percent} quantile of the random compliance, as defined in \eqref{eq:quant_def}. Feasibility is enforced probabilistically according to $\P_{\Omega}[c(\inp+\rnddsn) \leq 0] \geq 0.99$. Evaluating the volume of the bracket is computationally cheap and thus brute force MC ($n=10^6$) is feasible for evaluation of the probabilistic constraint.

% Simulation
We use the finite element method to compute the compliance for a given design point and fixed load conditions. To this end the domain $\mathcal{B}$ is discretized with triangular elements with linear shape functions (cf.~Figure \ref{fig:BracketModel}). This leads to a discretized system with around $7500$ degrees of freedom, depending on the design. Thus each data point for constructing/refining the mono-level surrogate requires assembly and solution of one linear finite-element problem.

\begin{figure}
	\centering
	\includegraphics[width=300pt]{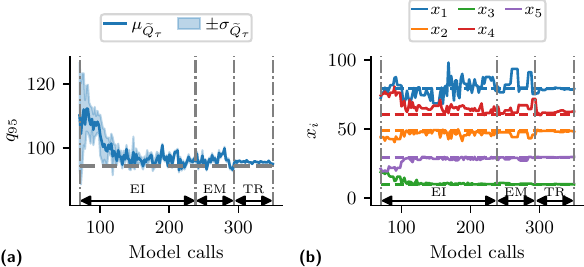}
	\caption{Evolution of the MLRO approach for structural optimization of the bracket model. The infill phases are denoted by the black arrows. \textbf{(a)} Mean and standard deviation of the current best quantile estimate. The global optimum is shown as gray dashed line. \textbf{(b)} Evolution of the current best design vector. The global optimum is shown as dashed line.}
	\label{fig:Bracket_evolution}
\end{figure}
\begin{figure}
	\centering
	\includegraphics[width=235pt]{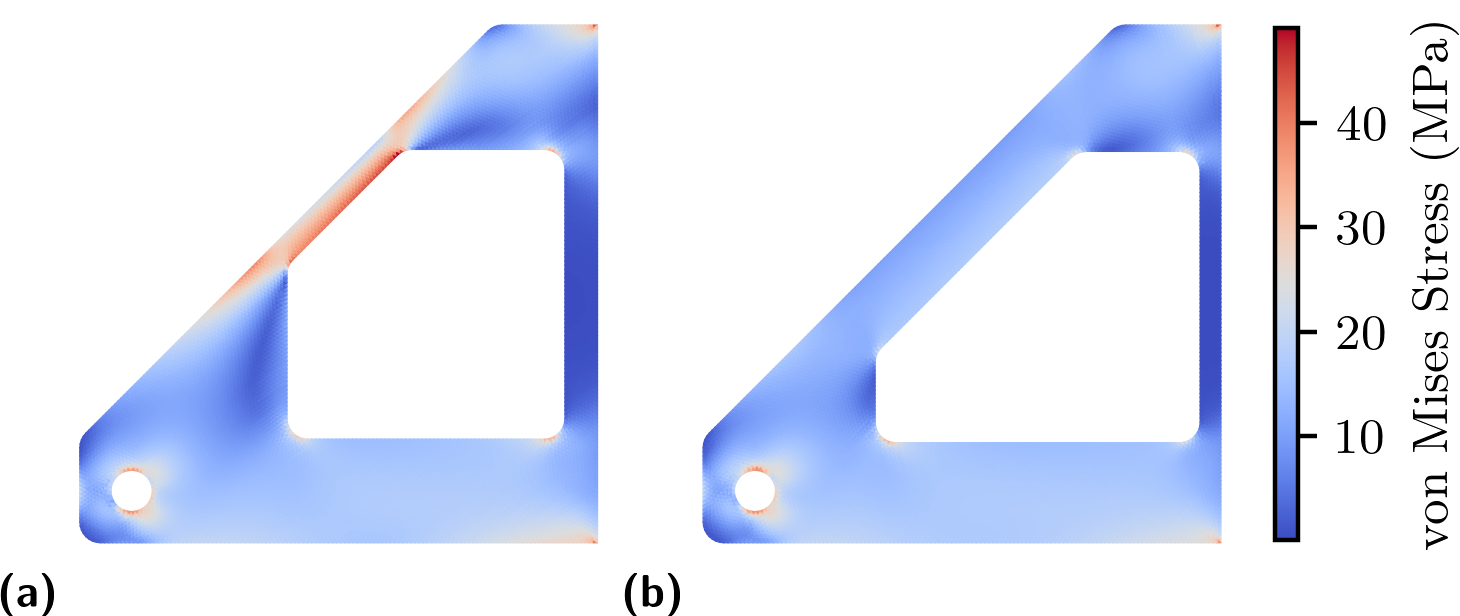}
	\caption{Mean von Mises stress field of the bracket structure subject to stochastic loading. \textbf{(a)}: Optimal design obtained from a deterministic design optimization. \textbf{(b)}: Optimal design obtained from a robust design optimization.}
	\label{fig:Bracket_stress}
\end{figure}
% Dicuss MLRO setup 
% Discuss the evolution 
Figure \ref{fig:Bracket_evolution}a shows the evolution of the best quantile estimate over model evaluations. At the beginning of the EI infill phase the quantile estimate has substantial epistemic uncertainty. This reflects the exploration behavior at that stage of the algorithm that requires surrogate-based quantile estimates in sparsely discovered regions. In the course of the optimization the algorithm focuses on promising regions and the uncertainty in the quantile estimate decreases and eventually the global optimum is approached. While the EI and EM phases are global stages, the TR phase is a local refinement. This becomes visible by the reduced oscillations in the TR phase. 

Figure \ref{fig:Bracket_evolution}b shows the evolution of the current best design vector. The reference optimum is shown by dashed lines and candidate samples are depicted as light markers. The variables $\inps_3$ and $\inps_5$ approach their target values early in the EI phase, already after \qty{30}{\percent} of that phase. Subsequent exploration primarily concerns $\inps_1$ and $\inps_4$ followed by $\inps_2$. 

% Discuss the accuracy
The final relative errors are \qty{2}{\percent} in the optimal design vector and \qty{0.7}{\percent} for the estimated quantile. The robustness measure is less sensitive to some design space directions, so a modest design error can still yield an almost optimal robustness measure. Next, we discuss the computational cost. The reference is obtained by a brute force genetic optimizer in the design space combined with a local adaptive surrogate to estimate the quantile and required around \qty{1e5} model evaluations. The MLRO approach only required \qty{3e2} model evaluations which is a saving of more than two orders of magnitude.

% Dicsuss the physical iterpretation
Finally, we interpret the robust design from a physical perspective. Figure \ref{fig:Bracket_stress} shows the mean von Mises stress fields under the uncertain loading conditions for two designs. The design depicted in Figure \ref{fig:Bracket_stress}a is obtained by a classical deterministic optimization under nominal and thus horizontal load conditions. The design in Figure \ref{fig:Bracket_stress}b is the one obtained by applying the MLRO approach. The optimal deterministic design experiences high stresses in the diagonal part of the structure when subject to varying loads. In contrast, the robust design does not encounter such stress concentration, because the loading uncertainty is incorporated during optimization.

\section{Discussion}
\label{sec:Discussion}

\paragraph{Performance.}
% Goal of the study
A key performance indicator of the mono-level approach is the number of required model evaluations to locate the robust optimum. To evaluate this performance indicator for the MLRO approach, we conduct a comparison with a state-of-the-art bi-level approach introduced in \citep{sabater2020efficient} using the implementation in \citep{bekemeyer2022data}. 

% Quick summary of the bi-level approach 
The bi-level approach combines Bayesian optimization in the design space with a separate surrogate-based uncertainty quantification procedure to estimate the quantile. Thus, a single GP is constructed and adaptively refined in the design space for the optimization task and at each training point a local GP is constructed to emulate the stochastic model behavior. This local GP for uncertainty quantification is again adaptively refined according to the target robustness measure. For further details of the approach we refer to \citep{sabater2020efficient}.

% Details on the setup fot the bi-level approach
For the outer GP we use standard computational budget allocation rules, $n_{\text{init}} = 10 d_{\inp}$, as used for the mono-level approach. For the sequential infill we use $n_{\text{seq}} = 2 n_{\text{init}}$. The sequential infill iterations are distributed as \qty{60}{\percent} EI, \qty{20}{\percent} EM, \qty{20}{\percent} TR as done for the mono-level approach.

The outer GP uses the converged quantile estimates obtained from the surrogate-based uncertainty quantification procedure as observations and hence accurate local estimates are crucial.
Therefore, the adaptive refinement of each local GP is continued  until the relative change in the estimated quantile falls below $10^{-3}$. To reduce the influence of fluctuations in the convergence history, the convergence criterion is based on the mean of the relative change between the current estimate $\tilde{q}_{(i)}$ and the previous two estimates, $\tilde{q}_{(i-1)}$ and $\tilde{q}_{(i-2)}$. The convergence criterion is expressed as
\begin{equation}
   \frac{1}{2} \sum_{k=1}^{2} \frac{ | \tilde{q}_{(i-k)} - \tilde{q}_{(i)} |}{| \tilde{q}_{(i)} |} < 10^{-3}.
\end{equation}

\begin{figure}
    \centering
    \includegraphics[width=290pt]{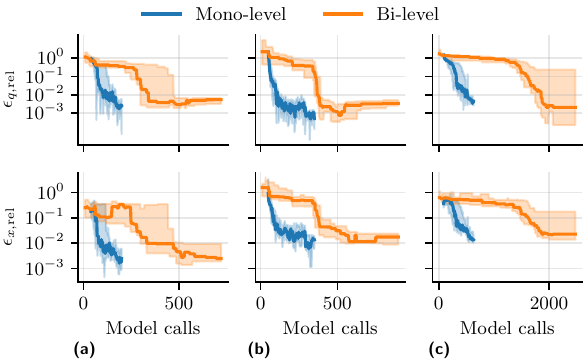}
    % Reproduce: 
    \caption{Comparison of the mono-level to a bi-level approach for \textbf{(a)} the Sabater test function (case $g_2$), \textbf{(b)} the Branin--Hoo test function and \textbf{(c)} the Hartmann-6 test function. The top row shows the relative error of the best quantile and the bottom row the relative error of the best design vector over model evaluations. All plots show the median over 10 runs with varying seeds as solid line and the \qty{80}{\percent} central probability region as shaded area.}
    \label{fig:MonoVsBilevel}
\end{figure}
% Discussion of results
Figure \ref{fig:MonoVsBilevel} compares the computational performance of the mono-level and the bi-level approach on the three analytical test functions introduced in Section \ref{ssec:analtest}. The mono-level approach requires fewer model evaluations to locate the global optimum than the bi-level approach. To quantify this statement we investigate the savings for different prescribed error thresholds $\epsilon_{\inp,\mathrm{rel}}^{(k)} = 0.01 + 0.01(k-1)$ for $ k=1,\ldots,10$. Define $N_{\mathrm{mono}}\big( \epsilon_{\inp,\mathrm{rel}}^{(k)} \big)$ and $N_{\mathrm{bi}} \big(\epsilon_{\inp,\mathrm{rel}}^{(k)} \big)$ as the numbers of model evaluations required for the median of the mono and bi-level runs, respectively, to reach the threshold $\epsilon_{\inp,\mathrm{rel}}^{(k)}$ and remain below it afterwards. The savings by the mono-level approach are then given as
\begin{equation}
\overline{f}
= 1 - 
\frac{1}{|\mathcal{K}|}
\sum_{k\in\mathcal{K}}
\frac{N_{\mathrm{mono}} \big( \epsilon_{\inp,\mathrm{rel}}^{(k)}\big)}
     {N_{\mathrm{bi}} \big(\epsilon_{\inp,\mathrm{rel}}^{(k)} \big)},
\end{equation}
where $\mathcal{K}$ contains only the thresholds reached by both methods. In these studies the savings $\overline{f}$ of the mono-level approach are in the range of \qtyrange{70}{77}{\percent}. 
For the considered test functions the MLRO approach leverages the strength of a global surrogate. It has the ability to work with rough quantile estimates and only refine the estimation in promising regions. However for higher dimensional problems with more complex behavior the bi-level approach may become more competitive, since it separates the optimization and uncertainty analysis task to two separate problems.

\paragraph{Flexibility.}
\label{ssec:flexibility}
Next, we investigate the flexibility of the MLRO approach. Suppose we have identified a robust design, however, it is now required to perform the optimization for a different robustness measure. Reusing the expensive model evaluations from the first optimization can substantially reduce the computational cost of the second optimization. 

Recall that the MLRO training dataset $\augDataset = \{(\auginp^{(i)}, \mdl(\auginp^{(i)})) \ \vert \ i = 1, \dots, n_{\text{total}}\}$ consists of labeled data in the augmented design space. These evaluations can therefore be reused for subsequent optimization tasks. The choice of the robustness measure however affects the sequential acquisition policies and thus the regions in the augmented design space that are explored in greater detail. In this study, we reuse the initial DOE data, which are independent of the robustness measure, as well as model evaluations collected during the EI stage. The latter are included because the EI stage includes exploration and is expected to provide useful information beyond the original optimization task. 

\begin{figure}
    \centering
    \includegraphics[width=190pt]{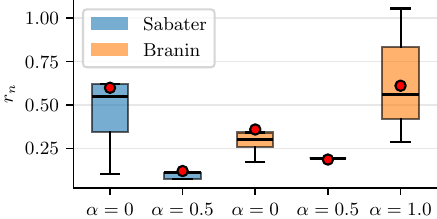}
    % Reproduce: 
    \caption{Reuse saving ratio in sequential model evaluations by reusing training data. Data generated during optimization of the $95\%$ quantile is reused for subsequent optimization of the mean--standard-deviation robustness measure. Results are shown for the Sabater and Branin--Hoo benchmarks and different values of $\alpha$.}
    \label{fig:Flexibility}
\end{figure}

To investigate the gain of information reuse within the MLRO approach we consider the Sabater and Branin-Hoo benchmarks from Section \ref{ssec:analtest}. Data generated during optimization of the \qty{95}{\percent} quantile are reused to optimize the mean--standard deviation objective \eqref{eq:mean_std_meas} for three different weights $\alpha \in \{0.0, 0.5, 1.0 \}$. We measure the benefit of information reuse using the evaluation ratio
\begin{equation}
    r^{(k)}_n = \frac{n^{(k)}_{\text{reuse}}}{\bar{n}_{\text{default}}}.
\end{equation}
Here $\bar{n}_{\text{default}}$ represents the average number of sequential model evaluations over multiple seed runs of a default optimization with no reused data. 
Complementary $n^{(k)}_{\text{reuse}}$ is the number of sequential training data for a single optimization run with information reuse. In both cases, we stop the count when the relative error in the design vector falls below a prescribed threshold. This threshold is set according to \qty{5}{\percent} for the Sabater test function and \qty{10}{\percent} for the Branin-Hoo test function. The reuse experiment is repeated for ten random seeds, and the box-plots of the evaluation ratio samples are shown in Figure~\ref{fig:Flexibility}.

Across all investigated cases, the average number of new sequential model evaluations relative to the default optimization is reduced. For $\alpha = 0.5$ the strongest reduction is observed: the reuse-based optimization requires only approximately \qtyrange{10}{20}{\percent} of the evaluations required by the default approach. In these cases the robust optimum is close to the optimum associated with the $95\%$ quantile, so its neighborhood has already been well explored.

For the Branin--Hoo case with $\alpha = 0$, information reuse reduces the required number of sequential evaluations by at least \qty{65}{\percent}, despite the change in the location of the robust optimum. For $\alpha = 1.0$ the reduction is lower and has a larger spread. This is because the global optimum lies in a flat valley that is in general difficult to optimize. Also, for the Sabater benchmark with $\alpha = 0$ the benefit strongly varies across different seeds. This behavior is caused by the similar robustness-measure values at two local minima, making this test-case in general difficult to solve. For $\alpha = 1$ the robust optimum is not unique and thus the case is omitted from this study. The different robustness measures considered in this section are visualized over the design space in Figures \ref{fig:sabater_transfer} and \ref{fig:branin_transfer} in Appendix~\ref{sec:plots_reuse}.  

\paragraph{Scalability.}
\begin{figure}
    \centering
    \includegraphics[width=230pt]{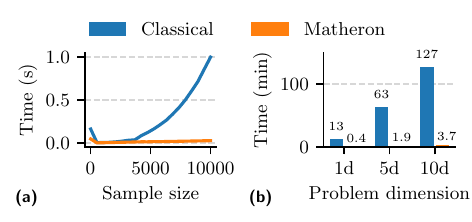}
    % Reproduce: 
    \caption{Computational cost for different posterior-sampling strategies. \textbf{(a)} Computing time for estimating the quantile as a function of the number $n$ of uncertainty-propagation test points, with $m=5000$ GP realizations. All experiments were run with PyTorch on a single NVIDIA A100 GPU \textbf{(b)} Computational time for a single optimization of the EI acquisition function for different design-space dimensions.}
    \label{fig:Scaling}
\end{figure}
Finally, we investigate the computational cost of the MLRO algorithm. A main computational bottleneck of the algorithm is the optimization of the design-space acquisition function, since it requires repeated computation of the robustness measure. In particular, evaluating the estimators in \eqref{eq:sampleEQ} and \eqref{eq:sampleStdQ} requires $m$ GP samples at $n$ MC sample points. Figure \ref{fig:Scaling}a displays the computational time required to estimate the quantile for different sample sizes $n$ with a number of GP realizations fixed at $m = 5000$. Classical GP posterior sampling exhibits the expected cubic scaling with $n$, resulting from the Cholesky factorization of the covariance matrix. In contrast, the efficient pathwise-sampling implementation based on Matheron's rule exhibits linear scaling as discussed in Section \ref{ssec:effSampling}.    

To relate these timings to the cost of acquisition function optimization, Figure \ref{fig:Scaling}b shows the approximate computational times required to optimize the EI acquisition function for different design space dimensions. We use differential evolution with a population size of $15 \din$ and a maximum of $50$ iterations. Other optimizers and settings are possible; the main observation is that for classical sampling the cost is in the order of hour while sampling via Matheron's rule reduces the cost to orders of minutes. 
In general the cost of acquisition function optimization should be assessed relative to the cost of evaluating the model $\mdl$ once. 
For complex computational models, it is feasible to spend a few minutes on optimizing the next infill point.

\section{Conclusion}
% Summary
% - Introduced a mono-level Surrogate based approach for robust optimization
In this paper we introduced a novel mono-level surrogate-based approach for robust optimization. The approach constructs a single GP surrogate over the augmented input space representing the design variables and the stochastic model parameters. The surrogate is then adaptively refined to stepwise reduce the epistemic uncertainty in the robustness measure and eventually locate the robust optimum. 

% Main results
% - efficient sampling approach allows scalability to 10 dimensional problems
% - by construction the approach allows reusabiliy for optimizaiton of different robustness measures
% - outperforms classical bi-level approach
We applied the proposed MLRO approach to three analytical benchmark functions of increasing complexity including non-linear, non-convex and multi-modal optimization landscapes. We also demonstrated the approach on a structural optimization benchmark considering uncertainty in geometric design variables and loading conditions.

A central challenge in robust optimization is the high computational cost, associated with the large number of required model evaluations for optimization and uncertainty quantification. The MLRO approach is designed to work in small-data settings and therefore enables optimization of complex simulation models. For the benchmark models considered, the MLRO approach required fewer model evaluations than a state-of-the-art bi-level surrogate approach. The core of the algorithm is a sampling-based uncertainty estimator of the robustness measure, enabling adaptive refinement of the mono-level surrogate. 
To improve the scalability of this estimator, we employed an efficient GP posterior-sampling procedure based on pathwise sampling and a finite-dimensional weight-space representation. As shown, this procedure significantly reduces the sampling cost compared to classical approaches. Finally, we demonstrated the flexibility of the MLRO approach by reusing existing model evaluations for the optimization of different robustness measures. This led to a further reduction of infill model evaluations ranging from \qty{20}{\percent} - \qty{90}{\percent} depending on the robustness measure and test function.

% Outlook
The current study considers problems with an augmented input space of moderate dimensionality, on the order of ten dimensions. Extending the MLRO approach to higher-dimensional problems is an important direction for future work. Dimensionality-reduction methods, such as partial least squares and active subspace methods, may provide a way to reduce the effective dimension of the surrogate model. Efficient optimization of the acquisition functions in higher dimensions poses another challenge. When sampling-based uncertainty propagation becomes computationally prohibitive, alternatives such as Thompson sampling may offer a more efficient strategy. Lastly, extending the approach to multi-objective robust optimization and potentially including expensive-to-evaluate inequality constraints are further important research directions.
% Closing
Overall, the proposed MLRO approach, provides a flexible framework for surrogate-based robust optimization of expensive simulation models, with particular strength in small-data settings.
\clearpage %%Remove this from your manuscript

\appendix

\section{Evolution plots for analytical benchmarks studies}
\label{sec:AddPlots}

\begin{figure}
    \centering
    \begin{tikzpicture}
        % Include the actual plot
        \node[anchor=south west] at (0.5,3.25) {
            \includegraphics[width=300pt]{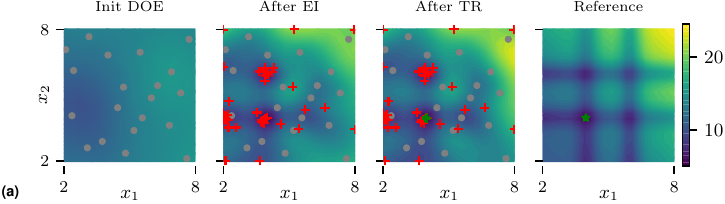}
        };
        \node[anchor=south west] at (0.6,0) {
            \includegraphics[width=300pt]{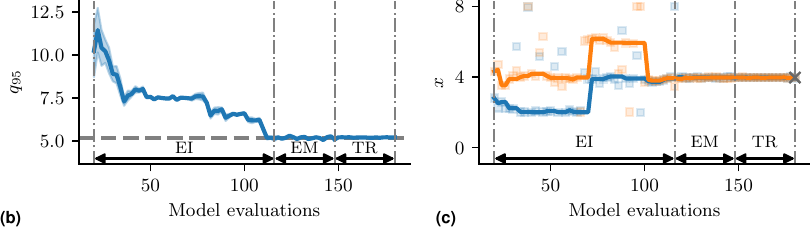}
        };
    \end{tikzpicture}
    % Reproduce: 
    \caption{Application of the mono-level approach to the Sabater test function with stochastic extension $g_2$. \textbf{(a)}: Snapshots of the mean surrogate-based quantile estimate after the successive infill phases. Markers represent the doe and infill samples. \textbf{(b)}:~Evolution of the robustness measure over model evaluations. The mean quantile prediction (blue), its standard deviation (light blue) and the optimum (gray) are presented. \textbf{(c)}: Evolution of the  design space samples over the model evaluations. Light markers denote candidate samples, while the solid line denotes the current optimum. The optimum is denoted as gray cross.}
    \label{fig:sabaterEvoQuad}
\end{figure}

\begin{figure}
    \centering
    \begin{tikzpicture}
        % Include the actual plot
        \node[anchor=south west] at (0.5,3.25) {
            \includegraphics[width=300pt]{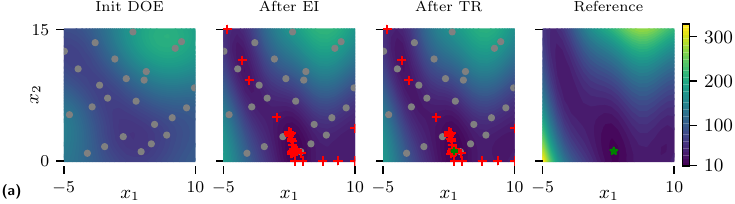}
        };
        \node[anchor=south west] at (0.6,0) {
            \includegraphics[width=300pt]{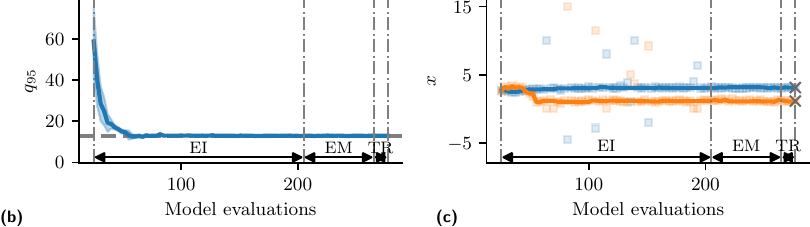}
        };
    \end{tikzpicture}
    % Reproduce: 
    \caption[Application of the mono-level approach to the Branin-Hoo test function]{Application of the mono-level approach to the augmented Branin-Hoo test function. \textbf{(a)}: Snapshots of the mean surrogate-based quantile estimate after the successive infill phases. \textbf{(b)}:~Evolution of the robustness measure over model evaluations. The mean quantile prediction, its standard deviation, and the error are presented. \textbf{(c)}: Evolution of the design space samples over the model evaluations. Light markers denote candidate samples, while the solid line denotes the current optimum.}
    \label{fig:braninEvo}
\end{figure}

This section presents additional results illustrating the evolution of the MLRO approach for the analytical benchmark functions introduced in Section \ref{ssec:analtest}. Figures \ref{fig:sabaterEvoQuad} and \ref{fig:braninEvo} show representative optimization runs for the Sabater benchmark with quadratic stochastic extension $g_2$ and the Branin-Hoo benchmark. Subfigure $\textsf{(a)}$ shows the mean quantile estimation after successive infill phases. The markers represent infill samples in the augmented design space projected onto $\dsnSpace$. In particular, the gray dots mark initial DOE samples, the red crosses EI infill samples and the green crosses EM and TR infill samples. Additionally the reference quantile prediction over the design space indicating the global optimum as green star is depicted on the right. Sub-figures $\textsf{(b)}$ and $\textsf{(c)}$ show the estimated quantile and the design vector, as a function of the number of model evaluations. 

For the Sabater test function, Figure \ref{fig:sabaterEvoQuad} shows that only a rough estimate of the quantile is available after the initial DOE phase. During the EI phase the algorithm explores the design space boundaries and regions near the local minima. As the number of model evaluations increases, the epistemic uncertainty in the quantile estimate decreases, as shown in Figure \ref{fig:sabaterEvoQuad}b.

For the Branin-Hoo test function, Figure \ref{fig:braninEvo} shows that the initial DOE is already sufficient to provide a reasonable approximation of the quantile in the design space. Consequently, the EI infill iterations mainly focus on exploring the valley of the reference optimum and eventually locate it. 

\begin{figure}
    \centering
    \includegraphics[width=300pt]{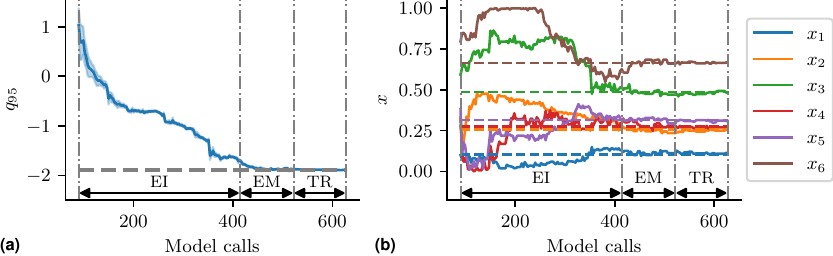}
    % Reproduce: 
    \caption[Application of the mono-level approach to the Branin-Hoo test function]{Evolution of a representative MLRO optimization run for the augmented Hartmann-6 test function.}
    \label{fig:hartmannEvo}
\end{figure}

Finally, Figure \ref{fig:hartmannEvo} plots a representative optimization run of the Hartmann-6 test function. The higher dimensional design space makes this problem more challenging thus more model evaluations are required to locate the reference optimum.

\section{DOE study for the Sabater $g_3$ benchmark}
\label{app:DOEstudySabat_g3}
\begin{figure}
    \centering
    \includegraphics[width=250pt]{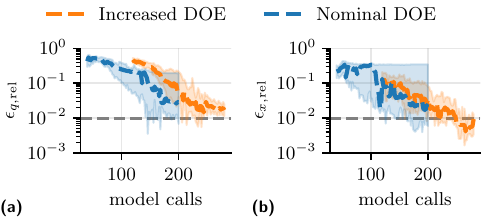}
    % Reproduce: 
    \caption{Comparison of the optimization evolution for the Sabater test function with stochastic extension $g_3$ for nominal and increased initial DOE. Nominal refers to the heuristics defined in Section \ref{ssec:Heuristics}. \textbf{(a)}: Evolution of the relative error in the optimal quantile. \textbf{(b)}: Evolution of the relative error in the best design vector.}
    \label{fig:sabater_g3_update}
\end{figure}

Five out of ten simulation runs did not locate the global optimum for the Sabater test function with strongly non-linear stochastic extension $g_3$. This indicates that the maximal computational budget and thus the amount of training data is not sufficient. Additional optimization runs with increased DOE are performed to check this assumption. Figure \ref{fig:sabater_g3_update} presents the median and the $80 \%$ central probability region for the nominal case and for increased number of initial DOE. Here, at each DOE point one additional infill iteration based on the variance criterion and one based on the U-function are performed. 
We observe that the relative error in the quantile as well as the relative error in the best design vector are reduced. While for the nominal case the relative error stays above $\qty{1}{\percent}$ the relative error now drops below $\qty{1}{\percent}$.

\section{Additional plots for information reuse studies}
\label{sec:plots_reuse}
\begin{figure}
    \centering
    \includegraphics[width=300pt]{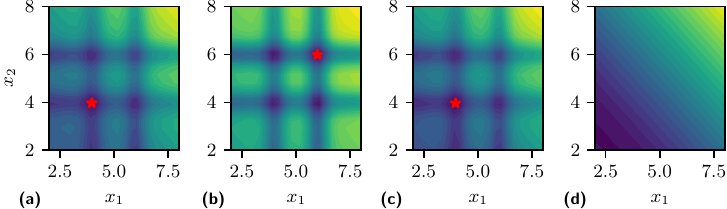}
    % Reproduce: 
    \caption{Visualization of different robustness measures over the design space for the Sabater test function with quadratic stochastic extension $g_2$. \textbf{(a)}: \qty{95}{\percent} quantile as robustness measure. The following use a weighted sum of mean and standard deviation as robustness measure with \textbf{(b)}: $\alpha = 0$, \textbf{(c)}: $\alpha = 0.5$, \textbf{(d)}: $\alpha = 1.0$. For this case there is no unique optimum.}
    \label{fig:sabater_transfer}
\end{figure}

\begin{figure}
    \centering
    \includegraphics[width=300pt]{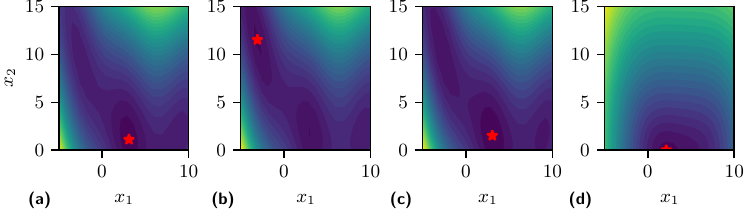}
    % Reproduce: 
    \caption{Visualization of different robustness measures over the design domain for the Branin-Hoo test function. \textbf{(a)}: \qty{95}{\percent} quantile as robustness measure. The following use a weighted sum of mean and standard deviation as robustness measure with \textbf{(b)}: $\alpha = 0$, \textbf{(c)}: $\alpha = 0.5$, \textbf{(d)}: $\alpha = 1.0$.}
    \label{fig:branin_transfer}
\end{figure}

This section provides additional plots for the study performed in Section \ref{ssec:flexibility} to assess the flexibility of the MLRO approach. Model evaluations obtained from optimizing the quantile are reused to optimize different robustness measures such as a weighted sum of mean and standard deviation. Figures \ref{fig:sabater_transfer} and \ref{fig:branin_transfer} visualize the corresponding robustness measures of the Sabater and Branin-Hoo benchmarks across $\inp$.

For the Sabater test function the \qty{95}{\percent} quantile robustness measure is plotted in Figure \ref{fig:sabater_transfer}a. Changing the robustness measure to the expectation leads to a new global optimum (Figure \ref{fig:sabater_transfer}b), while optimizing a weighted sum of mean and standard deviation with $\alpha = 0.5$ only results in a marginal change of the optimum (Figure \ref{fig:sabater_transfer}c). Taking the standard deviation as robustness measure leads to Figure \ref{fig:sabater_transfer}d.

For the Branin-Hoo test function the robust optimum changes with respect to the chosen robustness measure as well. The strongest change is observed when optimizing the expectation (Figure \ref{fig:branin_transfer}b). Optimizing the standard deviation or a weighted sum of mean and standard deviation with $\alpha = 0.5$ only results in a small change of the optimal design. 

\section*{Declaration of generative AI and AI-assisted technologies in the manuscript preparation process.}
During the preparation of this work the authors used ChatGPT5.6 and Claude Opus 5.5 for proofreading and language editing. After using this tool/service, the authors reviewed and edited the content as needed and take full responsibility for the content of the published article.

\section*{Acknowledgements}
We would like to acknowledge the funding by the Deutsche Forschungsgemeinschaft (DFG, German Research Foundation) under Germany’s Excellence Strategy—EXC 2163/1—Sustainable and Energy Efficient Aviation—Project-ID 390881007.

% To print the credit authorship contribution details
\printcredits

%% Loading bibliography style file
%\bibliographystyle{model1-num-names}
\bibliographystyle{cas-model2-names}

% Loading bibliography database
\bibliography{cas-refs}

\end{document}